\documentclass{aa}
\usepackage{txfonts} 
\usepackage{epsfig}
\usepackage{hangcaption}
\usepackage{graphicx}
\usepackage{rotating}
\usepackage{natbib}
\usepackage{array}              
\usepackage{dcolumn}
\usepackage{longtable}
\bibpunct{(}{)}{;}{a}{}{,}
\newcommand{\ns} {$\!\!\!\!\!$}
\newcommand{\mro} {\mathrm}
\newcommand{\degrees} {^\circ}

\begin{document}
\title{The geometry of \object{PSR~B0031$-$07}} \author{J.M. Smits \inst{1} \and D.
  Mitra \inst{2} \and B.W. Stappers \inst{3,4} \and J. Kuijpers \inst{1} \and
  P. Weltevrede \inst{4} \and A. Jessner \inst{5} \and Y. Gupta \inst{2}}
\offprints{J.M. Smits \email{Roy.Smits@\allowbreak manchester.\allowbreak ac.uk}}
\institute{Department of Astrophysics, Radboud University, Nijmegen, The
  Netherlands \and National Center for Radio Astrophysics, Pune, India \and
  ASTRON, Dwingeloo, The Netherlands \and Astronomical Institute `Anton
  Pannekoek', Amsterdam, The Netherlands \and Max-Planck-Insitut f\"ur
  Radioastronomy, Bonn} 
\date{Received <date> / Accepted <date>}

\abstract
{\object{PSR~B0031$-$07} is well known to exhibit three different modes of
  drifting sub-pulses (mode A, B and C). It has recently been shown that in a
  multifrequency observation, consisting of 2\,700 pulses, all driftmodes were
  visible at low frequencies, while at 4.85\,GHz only mode-A drift or
  non-drifting emission was detected. This suggests that modes A and B are
  emitted in sub-beams, rotating at a fixed distance from the magnetic axis,
  with the mode-B sub-beams being closer to the magnetic axis than the mode-A
  sub-beams. Diffuse emission between the sub-beams can account for the
  non-drifting emission.}
{Using the results of an analysis of simultaneous multifrequency
  observations of \object{PSR~B0031$-$07}, we set out to construct a
  geometrical model that includes emission from both sub-beams and diffuse
  emission and describes the regions of the radio emission of \object{PSR~B0031$-$07}
  at each emission frequency for driftmodes A and B.}  
{Based on the vertical
  spacing between driftbands, we have determined the driftmode of each
  sequence of drift. To restrict the model, we calculated average polarisation
  and intensity characteristics for each driftmode and at each frequency.}
{The model reproduces the observed polarisation and intensity characteristics,
  suggesting that diffuse emission plays an important role in the emission
  properties of \object{PSR~B0031$-$07}. The model further suggests that the emission
  heights of this pulsar range from a few kilometers to a little over 10
  kilometers above the pulsar surface. We also find that the relationships
  between height and frequency of emission that follow from curvature
  radiation and from plasma-frequency emission could not be used to reproduce
  the observed frequency dependence of the width of the average intensity
  profiles.}  
{} 
 \maketitle

\section {Introduction}
Pulsar \object{B0031$-$07} is well known for its three modes of drifting sub-pulses.
They are called mode A, B and C and are characterised by their values for
$P_3$ of 12, 7 and 4 times the pulsar period, respectively \citep{Huguenin70}.
This pulsar has been thoroughly studied at low observing frequencies
\citep{Huguenin70, Krishnamohan80, Wright81, Vivekanand95, Vivekanand97,
  Vivekanand99, Joshi00}, but only rarely at an observing frequency above
1\,GHz \citep{Wright81.3, Kuzmin86, Izvekova93}.  Recently, \citet{Smits05}
analysed simultaneous multifrequency observations from both the Westerbork
Synthesis Radio Telescope and the Effelsberg Radio Telescope and detected all
three drift modes at 325\,MHz, but only detected drift mode A at 4.85\,GHz.
The pulses that were classified as mode B or C at low frequency only showed
non-drifting emission at high frequency. On the basis of their findings, they
suggest a geometrical model where modes A and B at a given frequency are
emitted in two concentric rings around the magnetic axis with mode B being
nested inside mode A. This nested configuration is preserved across frequency
with the higher frequency arising closer to the stellar surface compared to
the lower one, consistent with the well known radius-to-frequency mapping operating in pulsars. Due to the rare
occurrence of mode C, they did not attempt to include this drift mode in their
model.

Here we analyse new multifrequency observations of \object{PSR~B0031$-$07},
obtained with the Giant Metrewave Radio Telescope, the Westerbork
Synthesis Radio Telescope and the Effelsberg Radio Telescope
simultaneously. In total, the observations contain 136\,000 pulses
spread over 7 different frequencies. From these observations we
attempt to restrict the geometry of this pulsar and create a model
that reproduces a great number of its observed characteristics.

In Section~\ref{sec:method} we explain how the observations have been
obtained, how the different modes of drift have been determined, and which
further analyses have been carried out. In Section~\ref{sec:results} we
present our results, which is followed by the modelling of the geometry of the
emission of \object{PSR~B0031$-$07} in Section~\ref{sec:modelling}. The discussion and
conclusions follow in Sections~\ref{sec:Discussion} and~\ref{sec:Conclusions},
respectively.

\begin{table}
  \caption{List of known parameters of \object{PSR B0031$-$07}. All values are
  from \citet{Taylor93}.}
  \begin{tabular}{ll}
  \hline
  \hline
  Parameter & Value\\
  \hline
  $P_1$ & 0.94295\,s\\
  $\dot{P}$ & $4.083\cdot10^{-16}$ \\
  DM & 10.89\,pc\,cm$^{-3}$ \\
  $S_{400}$ & 95\,mJy \\
  $S_{1400}$ & 11\,mJy \\
  $B_{surf}$ & $6.31\cdot10^{11}$\,G \\
  $\dot{E}$ & $1.9\cdot10^{31}$\,erg/s\\
  \hline
  \end{tabular}
  \label{tab:parameters}
\end{table}

\section{Method}
\label{sec:method}

\subsection{Definitions}
To describe the observational drift of sub-pulses we use three
parameters, which are defined as follows: $P_3$ is the spacing, at the
same pulse phase, between drift bands in units of pulsar periods
($P_1$); this is the ``vertical'' spacing when the individual radio
profiles obtained during one stellar rotation are plotted one above
the other (stacked). $P_2$ is the interval between successive
sub-pulses within the same pulse, given in degrees, and $\Delta\phi$,
the sub-pulse phase drift, is the fraction of pulse period over which a sub-pulse
drifts in one pulse period, given in $\degrees$/$P_1$. Note that
$P_2=P_3\times\Delta\phi$. 

These parameters are often thought to be associated with beams of
emission (sub-beams), rotating at a fixed distance around the magnetic
axis. When each sub-pulse that is seen to be drifting in consecutive
pulses (which is observed as a drift band) is due to emission from an
individual sub-beam, then $P_3$ is the rotation time of this
configuration divided by the number of sub-beams. However, when the
sub-beams rotate fast enough, it becomes possible that multiple
sub-beams contribute to one drift band. This is called aliasing and is
hard to detect. The observed $P_3$ is then no longer the rotation time
divided by the number of sub-beams, but depends on the degree of aliasing.

\subsection{Observations}
\label{sec:observations}
Here we present observations of \object{PSR~B0031$-$07}, which were obtained with the
Giant Metrewave Radio Telescope (GMRT), the Westerbork Synthesis Radio
Telescope (WSRT) and the Effelsberg Radio Telescope (EFF). They comprise 7
different frequencies. A great part of the observations are simultaneous at
different frequencies and include 4 hours in which radio emission from
\object{PSR~B0031$-$07} is observed at 5 frequencies simultaneously.

The 150, 243, 325, 610 and 1167 MHz observations were conducted using
the GMRT located in Pune, India. The GMRT is a multi-element aperture
synthesis telescope \citep{Swarup91} consisting of 30 antennas
distributed over a 25-km diameter area which can be configured as a
single dish \citep{Gupta00}. The antennas each have a gain of
0.3\,K/Jy. The system temperature for the different frequencies are
$T_{sys}^{157}$ = 482\,K, $T_{sys}^{243}$ = 177\,K, $T_{sys}^{325}$ =
108\,K, $T_{sys}^{607}$ = 92\,K and $T_{sys}^{1167}$ = 76\,K. The
signals from the antennas at 150, 243, 325 and 610\,MHz have circular
polarisation, while at 1167\,MHz the signals are linearly polarised.
At any frequency, orthogonally polarised complex voltages arrive
at the sampler from each of the antennas.  The voltage signals are
subsequently sampled at the Nyquist rate and processed through a
digital receiver system consisting of a correlator, the GMRT array
combiner (GAC) and a pulsar back-end for GAC. The signals selected by
the user are added in phase and fed to the pulsar back-end. The pulsar
back-end computes both the auto- and cross-polarised power, which was
then recorded at a sampling rate of 0.512\,ms. We have carried out
simultaneous observation at 243, 607 and 1167\,MHz by splitting the
whole array in to 3 different sub-arrays. We have used a scheme
wherein the digital sub-array combiner, which does the incoherent
addition of the multi-channel baseband data from different antennas,
is programmed to blank the data for a selected set of frequency
channels for antennas from a given sub-array. This is akin to setting
non-overlapping filter banks for each sub-array and allows the 16\,MHz
bandwidth to be divided between the different radio frequency bands of
observations, without any overlap between signals from the different
bands, while still preserving the time alignment of the data from the
different frequency bands. The 243 and 610\,MHz observations involved
polarimetry and the observations have been corrected for Faraday
rotation, dispersion and parallactic angle. Variations in the
rotation measure, due to intrinsic fluctuations of the ionosphere
during the length of one observation causes a deviation in the
position angle of less than 5$\degrees$ at an observation frequency of
243\,MHz. Towards higher frequencies, this deviation decreases with
$\lambda^2$. To correct for an instrumental polarisation effect, the
polarisation calibrator PSR B1929+10 was observed at several
parallactic angles, and was used to calibrate the data using the
technique described by \citet{Mitra05}. In most cases the corruption
due to leakage was found to be small, and could be corrected up to an
accuracy of 5\%.

The WSRT observations were made at a frequency of 840\,MHz with a bandwidth of
80\,MHz and recorded using the pulsar back-end, PuMa~\citep{Voute02}. The
signals from 14 telescopes were added with appropriate delays resulting in a
gain of 1.2\,K/Jy. The system temperature at this frequency is around
$T_{sys}$=150\,K. The Effelsberg observations were made at a frequency of
4.85\,GHz and a bandwidth of 500\,MHz. The gain is 1.5\,K/Jy and the system
temperature at this frequency is $T_{sys}$=27\,K. The polarisation calibration
is described in chapter 4 of \citet{Hoensbroech99}. The WSRT observations have
been corrected for Faraday rotation, dispersion and for any instrumental
polarisation effects using a procedure described in the Appendix of
\citet{Edwards04.2}.  

A 50-Hz signal present in the Effelsberg observation has been removed by
Fourier transforming the entire sequence, removing the 50\,Hz peak and Fourier
transforming back.  Table~\ref{tab:observations} lists the details of all the
observations. All the observations were aligned by correlating long sequences
of pulses with pulses from the observation at 607\,MHz that were obtained
simultaneously. This procedure removes any delay in arrival time between pulses at different frequencies due to retardation and aberration. 
The accuracy of this alignment is within 1\,ms.

\begin{table*}
  \caption{Details of the observations of \object{PSR~B0031$-$07}}
  \begin{tabular}{lllllll}
  \hline
  \hline
  Date       & Start time (UT)     & Telescope & Freq. & Time res. & Bandwidth & Number of \\    
             & (s after midnight) &           & (MHz)     & (ms) & (MHz)       & Pulses \\
  \hline
  02-09-2004 & 62616              & GMRT      & 325       & 0.512  & 16    & 2678 \\
  03-09-2004 & 69092              & GMRT      & 243       & 0.512  & 6.25  & 14607 \\
  03-09-2004 & 69092              & GMRT      & 607       & 0.512  & 9.75  & 14607 \\
  03-09-2004 & 72150              & WSRT      & 840       & 0.8192 & 80    & 16140 \\
  03-09-2004 & 75847              & EFF       & 4\,850    & 0.9206 & 500   & 31042 \\
  03-09-2004 & 80087              & GMRT      & 1\,167    & 1.024  & 16    & 10145 \\
  04-09-2004 & 00554              & GMRT      & 243       & 1.024  & 6.25  & 4409 \\
  04-09-2004 & 00554              & GMRT      & 607       & 1.024  & 9.75  & 4409 \\
  07-09-2004 & 69522              & GMRT      & 607       & 0.512  & 9.75  & 4718 \\
  07-09-2004 & 71441              & WSRT      & 840       & 0.8192 & 80    & 16781 \\
  07-09-2004 & 73690              & GMRT      & 243       & 0.512  & 6.25  & 14394 \\
  24-08-2005 & 43212              & GMRT      & 157       & 0.512  & 16    & 2229 \\
  \hline
  \end{tabular}
  \label{tab:observations}
\end{table*}

\subsection{Finding the drift sequences}
\citet{Smits05} have shown that for at least one of the drift modes the value
for $P_3$ is the same for both high and low frequency observations. This is
consistent with the concept that the emission of pulsars originates from
sub-beams of particles that rotate around the magnetic axis due to the
force-free motion of the particles in the strong magnetic field. Here, we have
investigated the values of $P_3$ for both mode A and mode B over a larger
frequency range, using the same method as described by \citet{Smits05}. From
the observations we chose 1\,500 consecutive pulses for which the sub-pulses
were visible in at least 4 frequencies. We then tested for each drift sequence
whether the value of $P_3$ was the same at each frequency. We can confirm that
in these 1\,500 pulses, within errors, $P_3$ remains constant for mode A over
the frequencies 243, 607, 840\,MHz and 4.85\,GHz. Furthermore, we unexpectedly
found multiple detections of mode-B drift at 4.85\,GHz, whereas
\citet{Smits05} found no mode-B drift in 2\,700 pulses at 4.85\,GHz, nor in
5\,350 pulses at an even lower frequency of 1.41\,GHz. Still, we could not
accurately measure $P_3$ at 4.85\,GHz when the pulsar was in mode B. We did
confirm that in this mode $P_3$ remains constant over the frequencies 243, 607
and 840 MHz. We then made the assumption that there is indeed one fundamental
period associated with the vertical spacing between drift bands, such as the
rotation speed of sub-beams around the magnetic axis, that is the same at each
frequency at each pulse phase. With this assumption, we determined the drift
mode of each pulse for all of the observations. This was achieved by visually
finding drift sequences at the frequency that showed this drift most clearly.
Whenever possible, these sequences were then inspected at the remaining
available frequencies to improve the exact beginning and end of the sequences.
Finally, $P_3$ was calculated for each sequence and classified as either mode
A, B, or C. Even though the time span of the observations is about 10 hours,
we only detected 4 short drift sequences with a mode-C drift, none of them
were detected at 4.85\,GHz. Because of the low number of pulses which were in
a mode-C drift, we limited the analysis to sequences of pulses which were in
mode-A or mode-B drift.

\subsection{Average polarisation profiles}
Once the drift class of all the sequences was known, all pulses in a
drift class were averaged together for each frequency separately. This
resulted in average mode-A and B polarisation profiles and
position angle sweeps at each frequency. We
also measured the widths of each of these profiles at 10\% of the peak
values.

\subsection{Frequency dependence of the fractional drift intensity}
\begin{figure}
\centering
\includegraphics[width=0.37\textwidth, angle=-90]{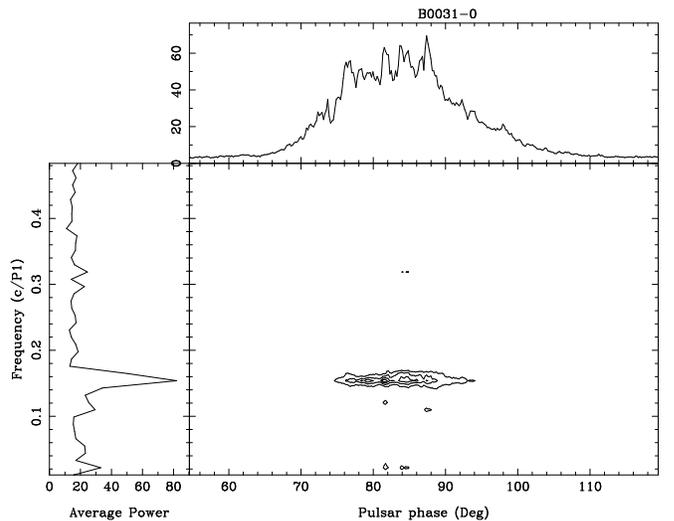} 
\caption{Contour plot of the LRFS of \object{PSR~B0031$-$07} at 0.325\,GHz as a function of pulse
  phase during a sequence of mode-B drift. The left panel shows the power
  spectrum averaged over phase. The upper panel shows the power
  averaged over frequency.}
\label{fig:LAFS}
\end{figure}

\citet{Smits05} have suggested that the mode-A and mode-B emission originate
from different regions of the magnetosphere. This means that the line of sight
intersection with the sub-beams is different for both drift modes (see their
figure 4 and 5). In particular, this causes the line of sight to miss the
center of the ring of mode-B sub-beams at high frequency and only cuts through
a diffuse component surrounding the sub-beams. As a result we do not see any
drift bands at high frequency whenever the low-frequency emission reveals a
mode-B drift. If the line of sight intersection with the sub-beams indeed
determines how clearly a drift pattern can be observed, then the drift pattern
can provide information about the geometry of the sub-beams. To measure how
far the line of sight intersection is away from the center of the sub-beams in
a drift sequence, we calculated the fractional drift intensity as follows.
First we took the amplitude at each pulse phase of each pulse in the sequence
and stacked them on top of each other to produce a 2D time series of the same
type as shown in Fig.~\ref{fig:grayscales}. We then calculated the absolute
values of the Fourier transform at each pulsar phase of this 2D time series.
The resulting Longitude-Resolved Fluctuation Spectrum (LRFS) was then averaged
over a pulse phase window containing the pulsar on pulse, giving a
Longitude-Averaged Fluctuation Spectrum (LAFS). A contour plot of a LRFS, as
well as the LAFS, is shown in Fig.~\ref{fig:LAFS}.  The drift intensity was
calculated by integrating the LAFS over a small frequency range containing the
reciprocal of the $P_3$ value of the sequence and then subtracting the
integration of the LAFS over a small frequency range containing no
periodicities. By dividing the drift intensity by the total intensity in the
LAFS of the pulsar signal we obtained the fractional drift intensity. This
fractional drift intensity was calculated for all sequences of drift at each
observed frequency. When the line of sight starts to miss the center of the
ring of sub-beams and only cut through diffuse emission, then the fractional
drift intensity should become low.

\subsection{Width of average drift profiles}
The geometrical model that will be constructed to describe the single pulse
emission from \object{PSR~B0031$-$07} consists of both drifting and non-drifting
emission. From the model we can calculate at each frequency and for each drift
mode the width of the average profile resulting from the drifting emission
only. In order to use this width to test how well the model describes the
observed emission, we needed to determine the width of the average profiles
resulting from the drifting emission only, in the observations. This was done
for each frequency as follows. For each drift sequence, we determined the
average drift profile by integrating the power in the LRFS at each pulse phase
over a small frequency range, containing the reciprocal of the $P_3$ value of
the sequence. For each drift mode these profiles were averaged together and
their widths were measured.

\subsection{Frequency dependence of $P_2$}
The change of $P_2$ with observation frequency reflects the change of the size
of the emission zone with respect to the total path of the line of sight
intersection at different heights above the pulsar surface. It therefore
limits the radius-to-frequency mapping that can be used for any model. $P_2$
is often calculated by averaging autocorrelations of single pulses, containing
more than one sub-pulse. For \object{PSR~B0031$-$07} this method fails at high
frequencies where there are no pulses containing more than one sub-pulse. We
therefore calculated $P_2$, for each drift sequence, as the product of $P_3$
and the sub-pulse phase drift ($\Delta\phi$) of that sequence. The phase drift
was measured by cross-correlating consecutive pulses of each drift sequence
and averaging the resulting cross-correlations. The peaks of the average
cross-correlations were fitted with a quadratic polynomial, of which the phase
shift of the maximum was taken as the phase drift. For each drift mode and
observation frequency the values for $P_2$ were averaged over many drift
sequences to obtain the frequency dependence of $P_2$ for each drift mode. The
values for $P_2$ are not affected by a possible aliasing.

\clearpage
\section{Results}
\label{sec:results}

\begin{figure*}
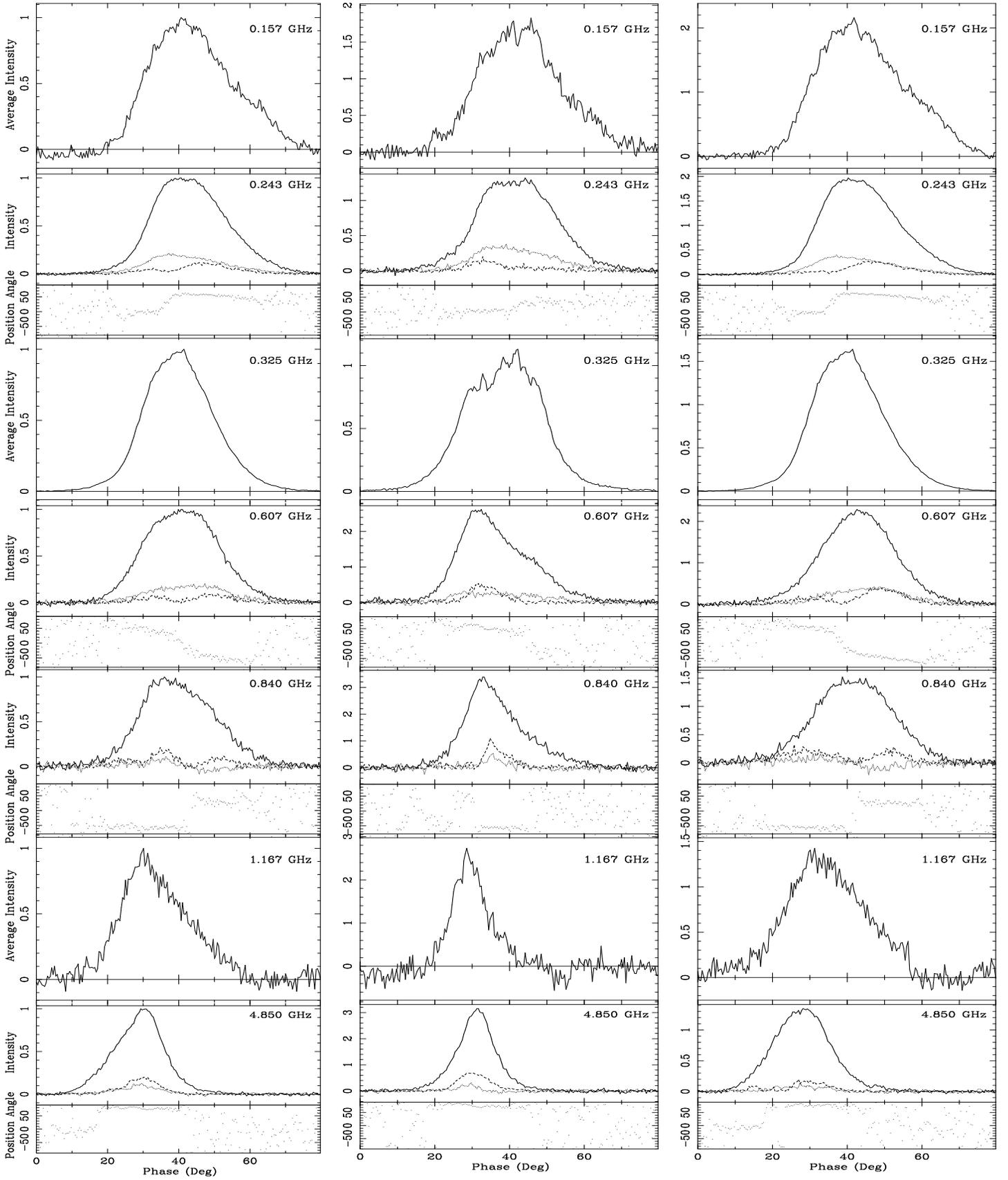

  \centering
  \begin{tabular}{ccc}
  \includegraphics[width=0.18\textwidth, height=0.33\textwidth, angle=-90]{eps/6292F01.ps} &
  \includegraphics[width=0.18\textwidth, height=0.33\textwidth, angle=-90]{eps/6292F02.ps} &
  \includegraphics[width=0.18\textwidth, height=0.33\textwidth, angle=-90]{eps/6292F03.ps} \\[-0.39cm]
  \includegraphics[width=0.185\textwidth, height=0.33\textwidth, angle=-90]{eps/6292F04.ps} &
  \includegraphics[width=0.18\textwidth, height=0.33\textwidth, angle=-90]{eps/6292F05.ps} &
  \includegraphics[width=0.18\textwidth, height=0.33\textwidth, angle=-90]{eps/6292F06.ps} \\[-0.44cm]
  \includegraphics[width=0.18\textwidth, height=0.33\textwidth, angle=-90]{eps/6292F07.ps} &
  \includegraphics[width=0.18\textwidth, height=0.33\textwidth, angle=-90]{eps/6292F08.ps} &
  \includegraphics[width=0.18\textwidth, height=0.33\textwidth, angle=-90]{eps/6292F09.ps} \\[-0.39cm]
  \includegraphics[width=0.185\textwidth, height=0.33\textwidth, angle=-90]{eps/6292F10.ps} &
  \includegraphics[width=0.18\textwidth, height=0.33\textwidth, angle=-90]{eps/6292F11.ps} &
  \includegraphics[width=0.18\textwidth, height=0.33\textwidth, angle=-90]{eps/6292F12.ps} \\[-0.43cm]
  \includegraphics[width=0.185\textwidth, height=0.33\textwidth, angle=-90]{eps/6292F13.ps} &
  \includegraphics[width=0.18\textwidth, height=0.33\textwidth, angle=-90]{eps/6292F14.ps} &
  \includegraphics[width=0.18\textwidth, height=0.33\textwidth, angle=-90]{eps/6292F15.ps} \\[-0.44cm]
  \includegraphics[width=0.181\textwidth, height=0.33\textwidth, angle=-90]{eps/6292F16.ps} &
  \includegraphics[width=0.18\textwidth, height=0.33\textwidth, angle=-90]{eps/6292F17.ps} &
  \includegraphics[width=0.18\textwidth, height=0.33\textwidth, angle=-90]{eps/6292F18.ps} \\[-0.38cm]
  \includegraphics[width=0.185\textwidth, height=0.33\textwidth, angle=-90]{eps/6292F19.ps} &
  \includegraphics[width=0.18\textwidth, height=0.33\textwidth, angle=-90]{eps/6292F20.ps} &
  \includegraphics[width=0.18\textwidth, height=0.33\textwidth, angle=-90]{eps/6292F21.ps} \\
  \end{tabular}
  \caption{Average polarisation profiles from all the
  observations. From left to right are the average profiles of all pulses, of
  pulses that are in mode A and of pulses that are in mode 
  B. The solid line is the intensity and the dashed and
  dotted lines are the linear and circular polarisation,
  respectively. Only the 243, 607, 840, 1167 and 4850\,MHz profiles
  were obtained simultaneously. All profiles are binned to the lowest
  time resolution of 1.024\,ms. The method for alignment is described in Section~\ref{sec:observations}}
  \label{fig:Profiles}
\end{figure*}

The (polarisation) profiles of \object{PSR B0031$-$07} for 7 frequencies are
shown in Fig.~\ref{fig:Profiles}. No polarisation was recorded for the
157, 325 and 1167\,MHz observations. The 157 and 325\,MHz observations
were not simultaneous with the other observations and were aligned to
have the center of the profiles at the same pulse phase as the center
of the 243 and 607\,MHz profiles, respectively. The widths at 10\% of
the maximum were measured for all the intensity profiles. They are
shown in Table~\ref{tab:widths}. The effects of dispersion are
negligible and have no significant contribution to the measured
width. The errors quoted here and elsewhere in the paper are 1-sigma
errors.

\begin{table}
  \centering
  \caption{The 10\%-widths of the average profiles from all pulses and
  from the drift modes A and B from the observations of \object{PSR~B0031$-$07} at 7
  different frequencies.}
  \begin{tabular}{D{.}{.}{4} r @{.} l c @{\ns$\pm$\ns} c r @{.} l  r @{.} l c @{\ns$\pm$\ns} c r @{.} l r @{.} l c @{\ns$\pm$\ns} c r @{.} l}
    \hline
    \hline
    \multicolumn{1}{l}{Frequency} & \multicolumn{6}{l}{width (deg)} & \multicolumn{6}{l}{width mode A} & \multicolumn{6}{l}{width mode B} \\
    \multicolumn{1}{l}{(MHz)}     & \multicolumn{6}{l}{\phantom{width mode A}} & \multicolumn{6}{l}{(deg)}        & \multicolumn{6}{l}{(deg)} \\
    \hline
    157       & 45&7 &&& 1&8 & \multicolumn{2}{l}{44} &&& \multicolumn{2}{l}{5} & 46&3 &&& 1&4  \\
    243       & 42&3 &&& 0&6 & 39&5 &&& 1&5 & 42&3 &&& 0&6  \\ 
    325       & 40&66&&& 0&15& 39&4 &&& 0&3 & 40&30&&& 0&12 \\ 
    607       & 39&8 &&& 1&1 & \multicolumn{2}{l}{32} &&& \multicolumn{2}{l}{4} & 39&3 &&& 0&9  \\
    840       & 40&9 &&& 0&6 & 32&2 &&& 0&8 & 43&0 &&& 0&9  \\
    1167      & \multicolumn{2}{l}{36} &&& \multicolumn{2}{l}{3} & \multicolumn{2}{l}{25} &&& \multicolumn{2}{l}{6} & \multicolumn{2}{l}{38} &&& \multicolumn{2}{l}{5} \\
    4850      & 33&7 &&& 0&8 & 23&6 &&& 0&6 & 32&8 &&& 0&9  \\
    \hline
  \end{tabular}
  \label{tab:widths}
\end{table}

\begin{figure*}
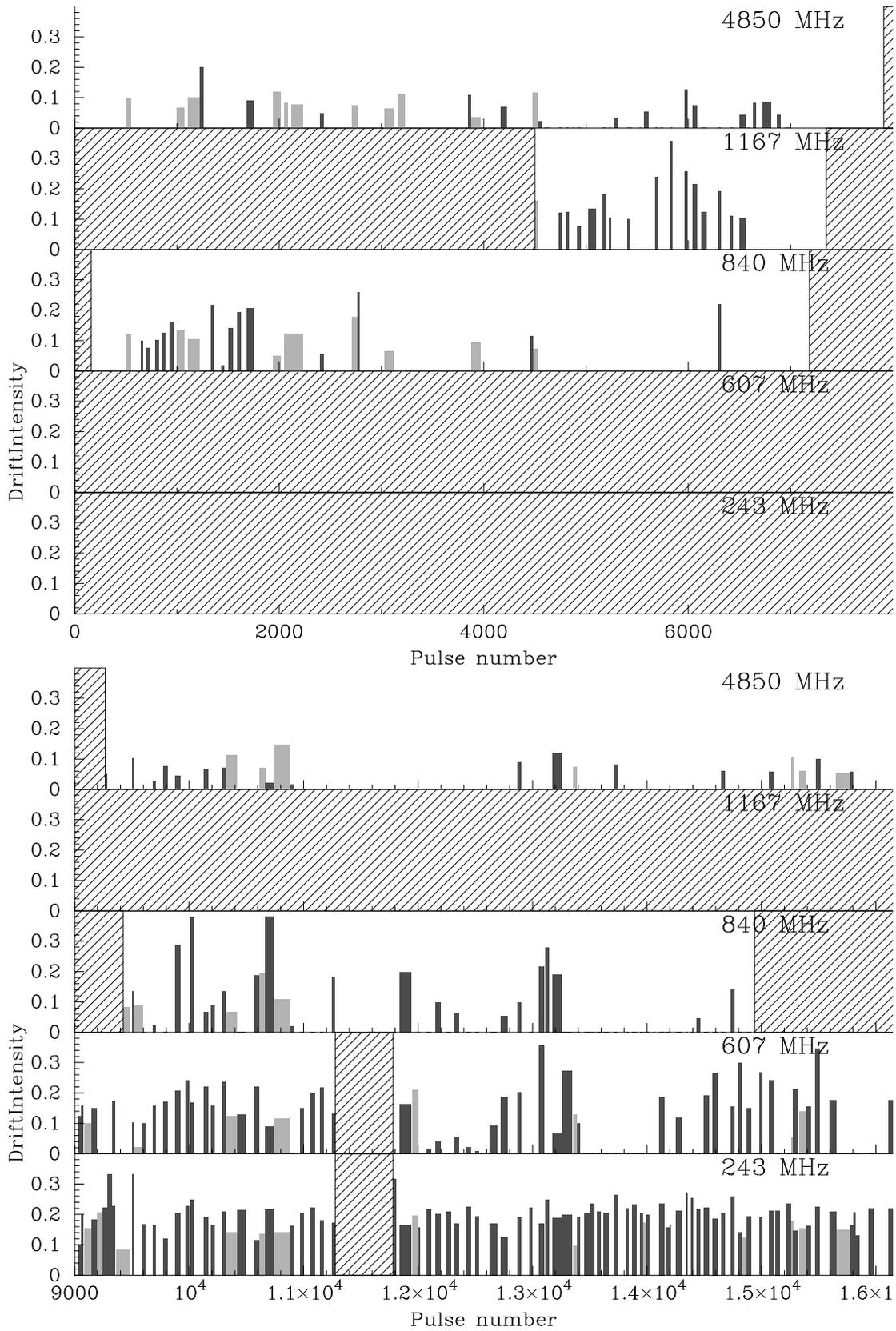

  \centering
  \includegraphics[angle=-90, width=0.80\textwidth]{eps/6292F22.eps}\\
  \includegraphics[angle=-90, width=0.80\textwidth]{eps/6292F23.eps}
  \caption{Fractional drift intensity in drift sequences at 5
    different frequencies. The gray columns represent sequences of
    pulses for which a drift intensity could be detected. Light-gray
    indicates mode-A drift, whereas dark-gray indicates mode-B
    drift. The hatched areas indicate regions during which no pulses
    were observed, or the quality of the observation was not good
    enough to determine the drift intensity. The lower graph is the
  continuation of the upper graph. The total duration is 4 hours
  and 15 minutes. }
  \label{fig:Driftranges}
\end{figure*}

Fig.~\ref{fig:Driftranges} shows the fractional drift intensity in
drift sequences for 5 different frequencies over a duration of 4 hours
and 15 minutes. Due to interference the drift intensity could not be
determined in some parts of the observations. These parts and parts
where no pulses were observed are marked in Fig.~\ref{fig:Driftranges}
as hatched areas. The average fractional drift intensities of drift
modes A and B for four of the frequencies are listed in
Table~\ref{tab:driftintensity}. The errors were estimated by
using the rms of the region in the LAFS that does not contain the
reciprocal of the $P_3$ value. The 1167\,MHz observation is not
listed because this observation does not contain enough drift
sequences to give a reliable average value. The widths of the average
drift profiles, i.e. of only the drifting component, for each
drift mode at each frequency are shown in
Table~\ref{tab:width_of_driftprofiles}. At 4.85\,GHz there was not
enough signal in the mode-B profile to measure a width.
\begin{table}
  \centering
  \caption{Average fractional drift intensity of drift mode A and B for
  four different frequencies.} 
  \begin{tabular}{l D{.}{.}{2} @{~$\pm$\ns} D{.}{.}{4} D{.}{.}{2} @{~$\pm$\ns} D{.}{.}{4}}
  \hline
  \hline
  Frequency & \multicolumn{2}{l}{Fractional}    & \multicolumn{2}{l}{Fractional}     \\
  (MHz)     & \multicolumn{2}{l}{drift intensity} & \multicolumn{2}{l}{drift intensity} \\
            & \multicolumn{2}{l}{in mode A}      & \multicolumn{2}{l}{in mode B}      \\
  \hline
  243       & 0.14 & 0.05    & 0.20 & 0.05\\
  607       & 0.07 & 0.02    & 0.11 & 0.03 \\
  840       & 0.07 & 0.02    & 0.04 & 0.02 \\
  4850      & 0.07 & 0.02    & 0.02 & 0.01 \\
  \hline
  \end{tabular}
  \label{tab:driftintensity}
\end{table}
\begin{table}
  \centering
  \caption{The 10\%-widths of the average drift profiles for drift mode A and B
  for 7 different frequencies. A `$-$' indicates that there was not
  enough signal to measure a width.}
  \begin{tabular}{l D{.}{.}{2} @{$\pm$\ns} D{.}{.}{2} D{.}{.}{2} @{$\pm$\ns} D{.}{.}{2} }
  \hline
  \hline
  Frequency & \multicolumn{2}{l}{10\%-width of} & \multicolumn{2}{l}{10\% width of} \\
  (MHz)     & \multicolumn{2}{l}{mode A (deg)}  & \multicolumn{2}{l}{mode B (deg)}  \\
  \hline
  157       & 28   & 2   & 35.0 & 1.5 \\
  243       & 27   & 5   & 35.2 & 0.9 \\
  325       & 33.4 & 0.4 & 35.6 & 0.4 \\
  607       & 27   & 3   & 30   & 4   \\
  840       & 12.2 & 0.9 & 10   & 4   \\
  1167      & 9    & 3   & 3.1  & 1.1 \\
  4850      & 16.4 & 0.9 & \multicolumn{1}{r@{$-$\ns}}{} & \\
  \hline
  \end{tabular}
  \label{tab:width_of_driftprofiles}
\end{table}
Table~\ref{tab:P2} lists the average values of $P_2$ for
drift modes A and B for 7 different observation frequencies. Note that
the observations at 157 and 325\,MHz were not part of the simultaneous
observations.

\begin{table}
  \centering
  \caption{The average values of $P_2$ for drift modes A and B
  for 7 different frequencies. A `$-$' indicates that there was not
  enough signal to measure $P_2$.}
  \begin{tabular}{l D{.}{.}{2} @{$\pm$\ns} D{.}{.}{2} D{.}{.}{2} @{$\pm$\ns} D{.}{.}{2} }
  \hline
  \hline
  Frequency & \multicolumn{2}{l}{$P_2$ of mode A} & \multicolumn{2}{l}{$P_2$ of mode B} \\
  (MHz)     & \multicolumn{2}{l}{(deg)}  & \multicolumn{2}{l}{(deg)}  \\
  \hline
  157       & 18.9 & 1.2 & 19.2 & 0.7 \\
  243       & 24   & 4   & 18.7 & 1.6 \\
  325       & 19.8 & 1.0 & 18.7 & 0.6 \\
  607       & 21   & 3   & 18.7 & 0.6 \\
  840       & 10   & 3   & 14.0 & 1.9 \\
  1167      & 14   & 5   & 12   & 3   \\
  4850      & 14   & 2   & \multicolumn{1}{r@{$-$\ns}}{} & \\
  \hline
  \end{tabular}
  \label{tab:P2}
\end{table}

\clearpage
\section{Modelling the geometry}
\label{sec:modelling}
Here, we set out to find a geometry that can reproduce a great number
of the observed features of this pulsar, which includes the
position angle sweep (without the orthogonal
mode jumps) and the frequency dependences of the width of the average
intensity profile, the width of the average drift profile, the
fractional drift intensity and $P_2$, for drift modes A and B.

\begin{figure}
  \centering
  \includegraphics[width=0.45\textwidth]{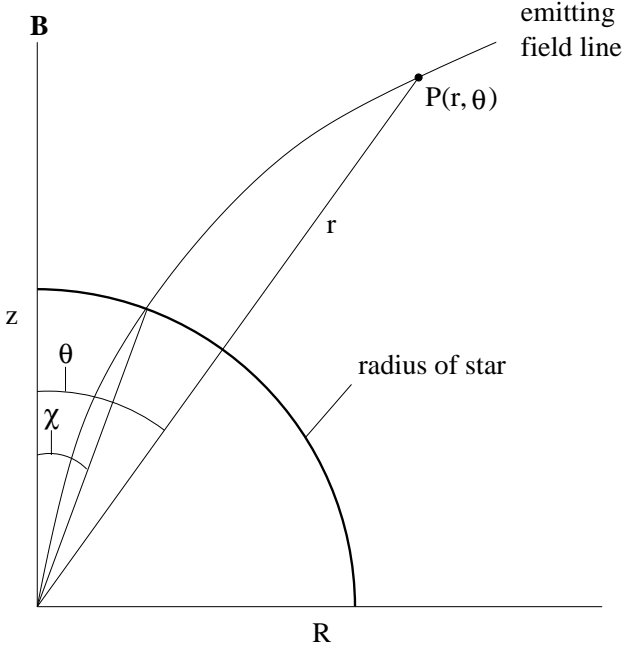}
  \caption{Sketch of a field line as a function of distance along the
  magnetic axis ($z$) and distance from the magnetic axis ($R$) near
  the pulsar, to show the definition of the angle $\chi$. Point $P$ on
  an emitting field line is defined by it's polar coordinates $r$ and
  $\theta$. $\chi$ is defined as the angle between the magnetic axis
  and the foot point of the field line on the surface of the star.}
  \label{fig:chi}
\end{figure}

We first try to limit the geometrical possibilities by fitting the
widths of the observed average profiles to three different models that
restrict the relationship between height and frequency of
emission. All models assume that the intensity of the radio emission
decreases with distance from the maximally emitting ring of field
lines as a Gaussian given by
\begin{equation}
I=\exp\left(-\frac12\left(\frac{\chi-\chi_0}{\chi_\mro{w}}\right)^2\right),
\end{equation}
where $I$ is the intensity, $\chi$ is the angle between the magnetic
axis and the foot point on the surface of the star of the field line
from which emission can be observed (see
Fig.~\ref{fig:chi}), $\chi_0$ is the angle between the
magnetic axis and the foot point on the surface of the star of the
maximally emitting field line and $\chi_\mro{w}$ is the angle between
the foot points on the surface of the star of the two field lines for
which the intensity has dropped to $e^{-\frac12}$ times the maximum
intensity.

In the first model, the curvature radiation model, we assume that the
frequency of emission is equal to the characteristic frequency for
curvature radiation, given by (\citep{Jackson99}, Eq. 14.81)
\begin{equation}
\omega = \frac{3}{2}\gamma^3\left(\frac{c}{r_\mro{cr}}\right),
\label{eq:curvature_frequency}
\end{equation}
where $\gamma$ is the Lorentz factor of the secondary plasma, $c$ is
the speed of light and $r_\mro{cr}$ is the radius of curvature at the
emission point. Following the derivation from \citet{Gangadhara04}, the dipolar geometry of the magnetic
field dictates how the radius of curvature is related to the distance
between the emission point and the center of the star, $r$, and the
polar angle to the magnetic axis, $\theta$, as
\begin{equation}
r_\mro{cr} = \frac{r}{\sin\theta}\frac{[5+3\cos(2\theta)]^\frac32}{3\sqrt2[3+\cos(2\theta)]}
\label{eq:altitude1}
\end{equation}
For angles smaller than 10$\degrees$ this can be linearized
within an error of 1.5\% to give
\begin{equation}
r_\mro{cr}=\frac43r/\theta.
\label{eq:altitude2}
\end{equation}

In the second model, the plasma-frequency model, we assume that the
radiation is emitted at the local relativistic plasma frequency, given
in SI by
\begin{equation}
\omega_\mro{rp}^2=\langle\gamma\rangle^{-1}\frac{e^2n}{\epsilon_0m_\mro{e}},
\label{eq:plasma_frequency}
\end{equation}
where $n$ is the sum of electron and positron density and $e$ and $m_\mro{e}$
are the electron charge and mass.  We further assume that there is a distinct
region around the magnetic axis where shots of particles can create a
secondary pair plasma with a fixed particle density which only depends on
altitude.  The frequency of occurrence of these shots is assumed to decrease
as a Gaussian with increasing distance from the maximum intensity field lines.
Thus, the intensity of the emission falls off with distance from the
maximally emitting field lines, while the particle density of the shots and
the frequency of emission remains constant at constant emission height. Since,
along the open field lines, the density of relativistically moving secondary
pair plasma falls off approximately with the cube of the distance to the
center of the star, in a dipolar field, on account of flux conservation, the
frequency drops with distance to the power $\frac32$. This leads to the
following relationship between the altitude and frequency of emission
\citep{Ruderman75}:
\begin{equation}
r=\left(\frac{\omega}{\omega_0}\right)^{-\frac23}
\label{eq:altitude3}
\end{equation}

Finally, we try a version of an empirical relationship, as given by \citet{Thorsett91}:
\begin{equation}
r=h_0\left(1+\left(\frac{\omega}{\omega_0}\right)^{-\frac23}\right),
\label{eq:altitude4}
\end{equation}
which is similar to the plasma-frequency model, but has an extra
distance parameter $h_0$ that allows for a minimum emission
height. For all three models we simulated the intensity profile at 7
different frequencies, corresponding to the 7 frequencies of the
observations, for different values of the parameters of the
models. These parameters and the limits within which they were varied,
are listed in Table~\ref{tab:Model_parameters}. $\gamma$ and
$\omega_0$ were varied logarithmically, the other parameters
linearly. We then calculated the widths at 10\% of the maximum. From
these widths and the widths from the mode-A and mode-B intensity
profiles from the observations we calculated the reduced chi-square
for each set of parameters for each drift mode. We then minimized
these chi-squares to fit the values for the parameters. Initial values
for $\alpha$ and $\beta$ were found by fitting the rotating vector
model from Radhakrishnan \& Cooke \citep{Radhakrishnan69} to the
position angle from all the observations for which polarisation was
recorded.  As can be seen in the panels of Fig.~\ref{fig:Profiles},
the position-angle sweep is very straight, apart from orthogonal mode
jumps which causes differences in the position-angle sweep at
different frequencies. After excluding the pulsar phases that contain
an orthogonal mode jump, each sweep can be reproduced with the
rotating vector model from Radhakrishnan \& Cooke, when $\alpha$ and
$\beta$ are kept equal and in the range of
0.1$\degrees$ to 6$\degrees$.

Once an initial relationship between the altitude and frequency of emission
was found, we proceeded to include both diffuse emission and emission from
rotating sub-beams to produce single pulses in two different drift modes for a
number of frequencies. The parameters of this model were then fitted to
reproduce the position angle sweep and the frequency dependences of the width
of the average intensity profile, the width of the average drift profile, the
fractional drift intensity and $P_2$, for drift modes A and B. Also, we used
the result from \citet{Smits05} that shows that for drift mode A at low frequencies, the
line of sight intersection just penetrates the maximally emitting field lines,
causing the observed double component in the upper-left panel of their Figure
8.

\subsection{Results of the Modelling}

\begin{table}
  \centering
  \caption{Parameters and their ranges of variation used to fit
  the curvature-radiation model (top), the
  plasma-frequency model (middle) and the empirical model
  (bottom).}
  \begin{tabular}{lll}
  \hline
  \hline
  Parameters for the & lower limit & upper limit \\
  curvature radiation &            &             \\
  model               &            &             \\
  \hline
  $\alpha$  & 0.1$\degrees$ & 6$\degrees$ \\ 
  $\beta$   & 0.1$\degrees$ & 6$\degrees$ \\ 
  $\gamma$  & 10          & 2000 \\
  $\chi_0$  & 0.3$\degrees$ & 0.92$\degrees$ \\
  $\chi_\mro{w}$  & 0.06$\degrees$  & 0.57$\degrees$ \\
  \hline
  \hline
  Parameters for the & lower limit & upper limit \\
  plasma-frequency   &             &             \\
  model              &             &             \\
  \hline
  $\alpha$  & 0.1$\degrees$ & 6$\degrees$ \\ 
  $\beta$   & 0.1$\degrees$ & 6$\degrees$ \\ 
  $\omega_0$& 0.01\,GHz     & 1\,000\,GHz \\
  $\chi_0$  & 0.3$\degrees$ & 0.92$\degrees$    \\
  $\chi_\mro{w}$  & 0.06$\degrees$ & 0.57$\degrees$ \\
  \hline
  \hline
  Parameters for the & lower limit & upper limit \\
  empirical model   &             &             \\
  \hline
  $\alpha$  & 0.1$\degrees$ & 6$\degrees$ \\ 
  $\beta$   & 0.1$\degrees$ & 6$\degrees$ \\ 
  $h_0$     & 10\,km      & 150\,km     \\
  $\omega_0$& 0.01\,GHz   & 1\,000\,GHz \\
  $\chi_0$  & 0.3$\degrees$ & 0.92$\degrees$    \\
  $\chi_\mro{w}$  & 0.06$\degrees$ & 0.57$\degrees$ \\
  \hline

  \end{tabular}
  \label{tab:Model_parameters}
\end{table}

For both the curvature-radiation model and the plasma-frequency model
we did not find any sets of values for the parameters to describe the
frequency dependence of the profile widths correctly. The best set of
parameters gave a reduced chi-square above 300 and above 15 for the
curvature-radiation and the plasma-frequency model, respectively. In
contrast, we found that the empirical model, which includes an extra
parameter, described the data with a reduced chi-square ranging from
0.7 to 3 for many different values for the parameters. Characteristic
to these sets of parameters were low values for $h_0$ and $\omega_0$,
suggesting a minimum emission height ranging from 1 to 50\,km above
the stellar surface and a maximum emission height ranging from 10 to
100\,km above the stellar surface, where the lowest values for the
chi-square always correspond to the lowest emission heights. We
did not allow emission heights below the stellar surface, but such
values would give low values for the reduced chi-square as well.  For
comparison, we tried to obtain emission heights by modelling the
emission for different values of the parameters $\alpha$, $\chi_0$ and
$\chi_\mro{w}$ ($\beta$ was kept equal to $\alpha$) and making no
assumptions about the relationship between height and frequency of
emission. For each frequency, the emission height was changed in such
a way that the width of the profile from the observation would be
equal to the width of the profile from the model. This always resulted
in small emission height differences between frequencies, varying from
a few kilometers to a few tens of kilometers. Also, for each set of
parameters it was possible to choose values for $h_0$ and $\omega_0$
for which the empirical model would fit well to the
relationship obtained between height and frequency of emission.

Based on these results we proceeded to make a more complex model, including
two drift modes of sub-pulse emission superposed on low-intensity diffuse
emission, assuming the empirical model for obtaining the emission height for
each frequency. The parameters of this model are listed in
Table~\ref{tab:Complete_model_parameters}, where $R_\star$ is the radius of
the pulsar, $D_\mro{w}$ is the angle between the foot points on the surface of
the star of the two field lines for which the intensity from the diffuse
emission has dropped to 60\% of the maximum intensity, $D_\mro{I}$ is the
intensity of the diffuse emission divided by the intensity of the sub-pulse
emission at the center field line, $N_\mro{sub}$ is the number of sub-pulses
and $P_3$ is the rotation time of the carousel divided by $N_\mro{sub}$. The
superscripts A and B refer to the value for that parameter in drift mode A or
B. Note that we only changed the values of $\chi_0$ and $P_3$ to describe a
change in drift mode. The rotation time of the carousel and the number of
sub-beams were derived from the observed vertical spacing between drift bands,
assuming that the observed sub-pulses are not aliased. If we are in fact
observing an alias, then the rotation time of the carousel as well as the
number of sub-pulses become less.  Table~\ref{tab:Complete_model_parameters}
also lists the values of these parameters that were obtained after fitting
them to describe the aforementioned features of the observations. The values
for $h_0$ and $\omega_0$ in Table~\ref{tab:Complete_model_parameters} lead to
emission heights ranging from 2.3\,km, at 4.85\,GHz, to 13.6\,km at 0.157\,GHz
above the stellar surface. Regardless of the relationship between height and
frequency of emission, these low emission heights are needed to obtain the
frequency dependence of $P_2$, which depends on the fractional change of
emission height. The value for $\chi_0^A$ corresponds to the last open field
line, which for a small value of $\alpha$ is given by $\chi_\mro{lof} = \sqrt
{R_\star 2\pi/P_1c}$.

\begin{figure}
  \centering
  \begin{tabular}{cc}
  \includegraphics[width=0.20\textwidth]{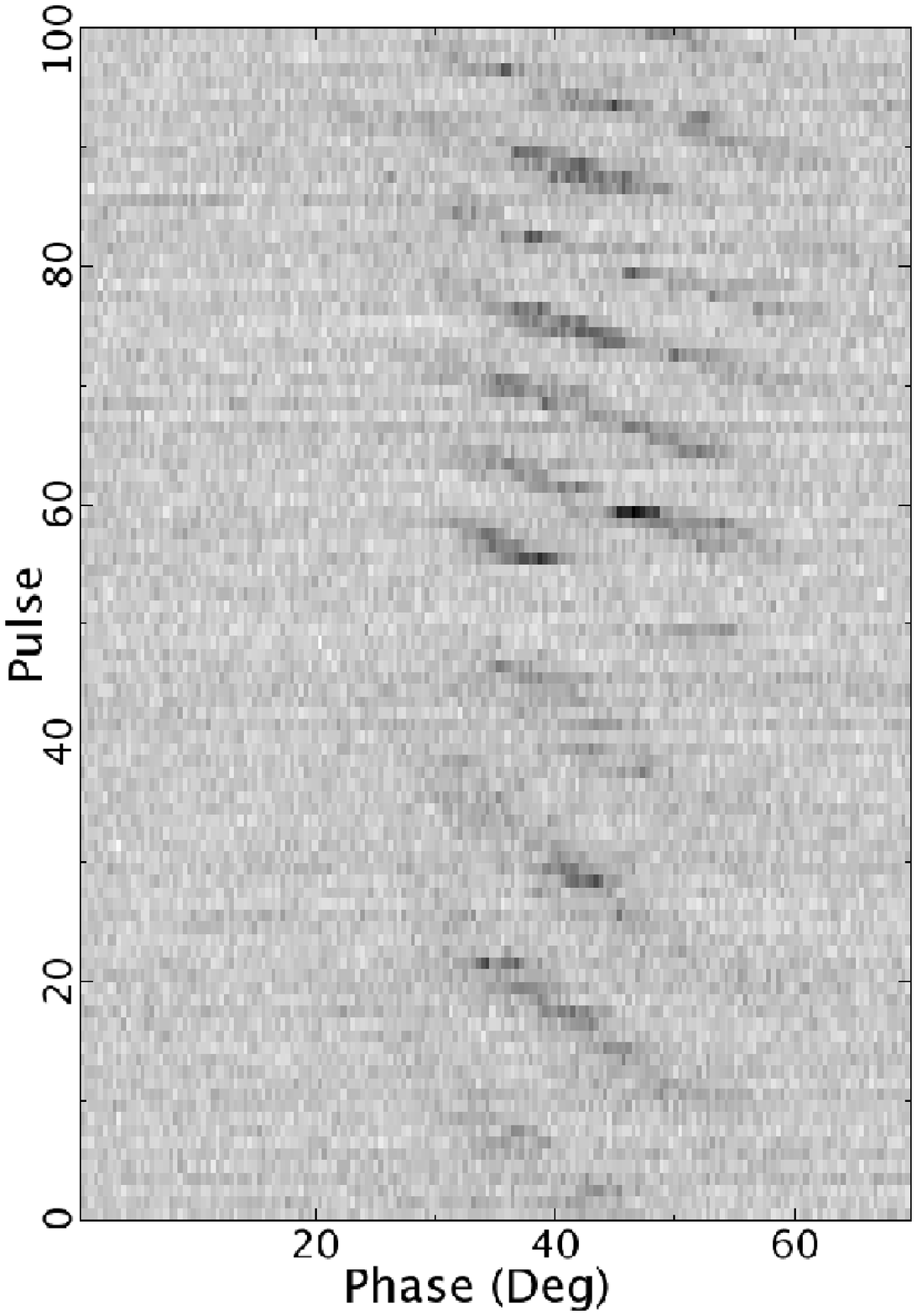} &
  \includegraphics[width=0.20\textwidth]{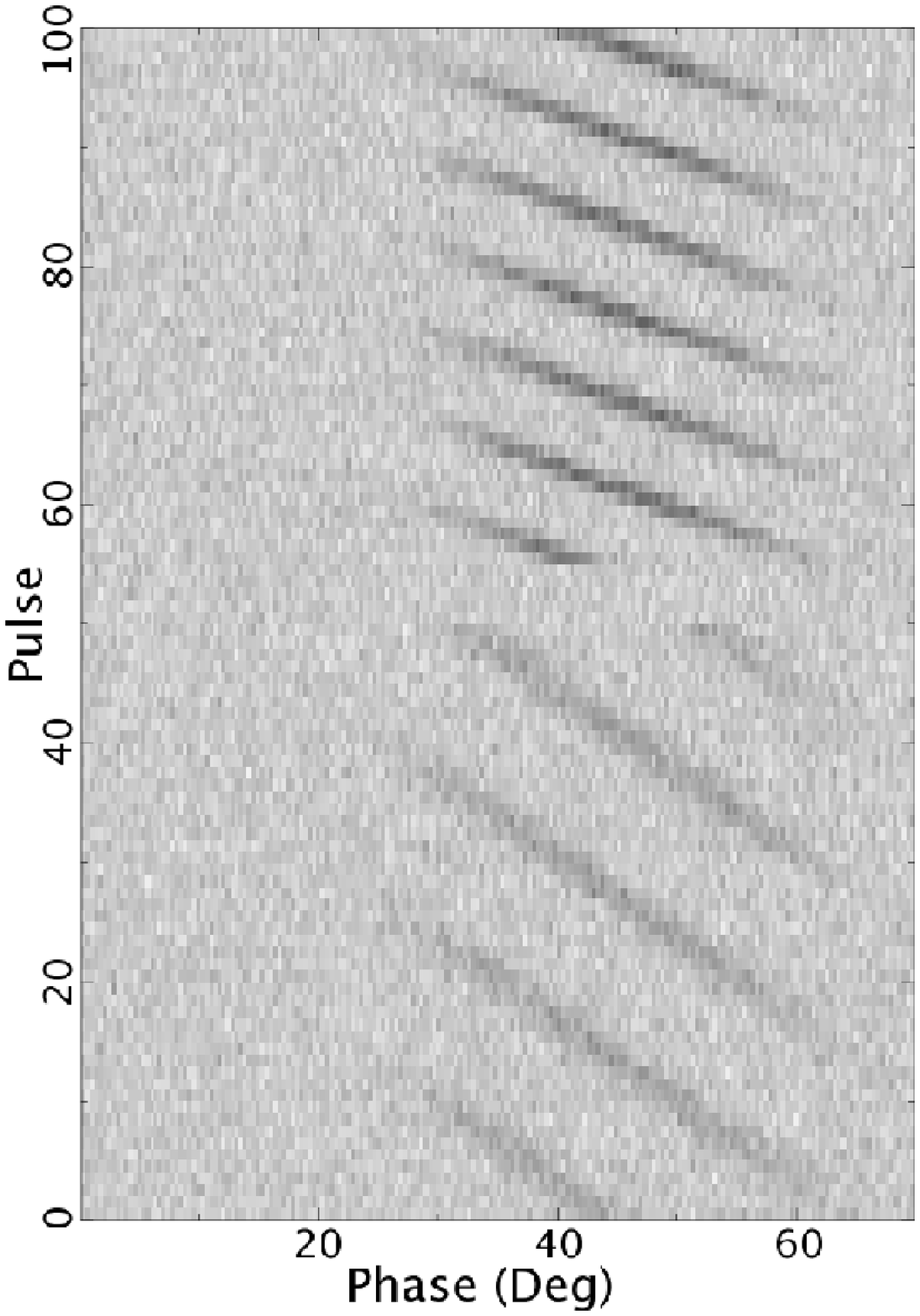} \\
  \includegraphics[width=0.20\textwidth]{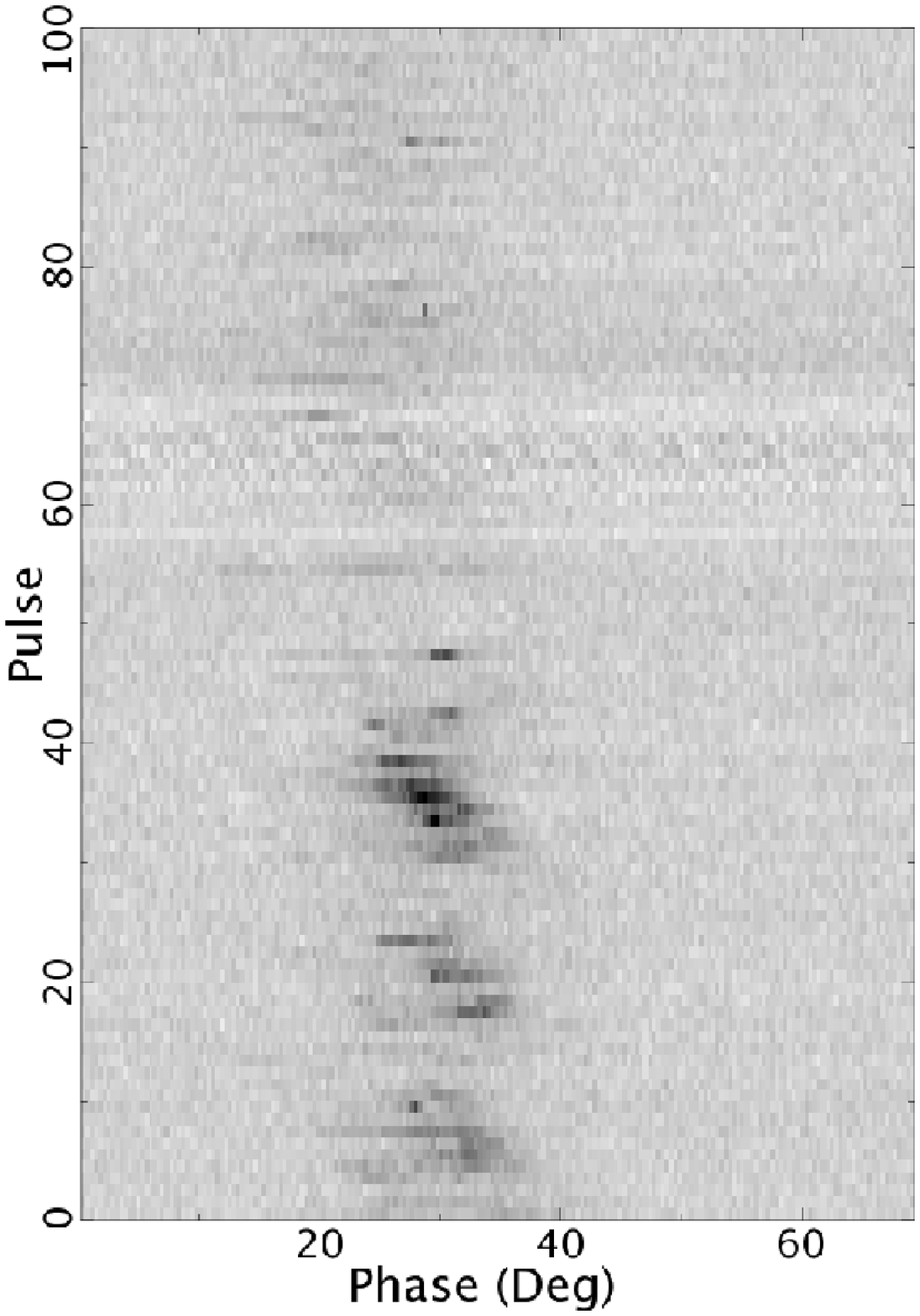} &
  \includegraphics[width=0.20\textwidth]{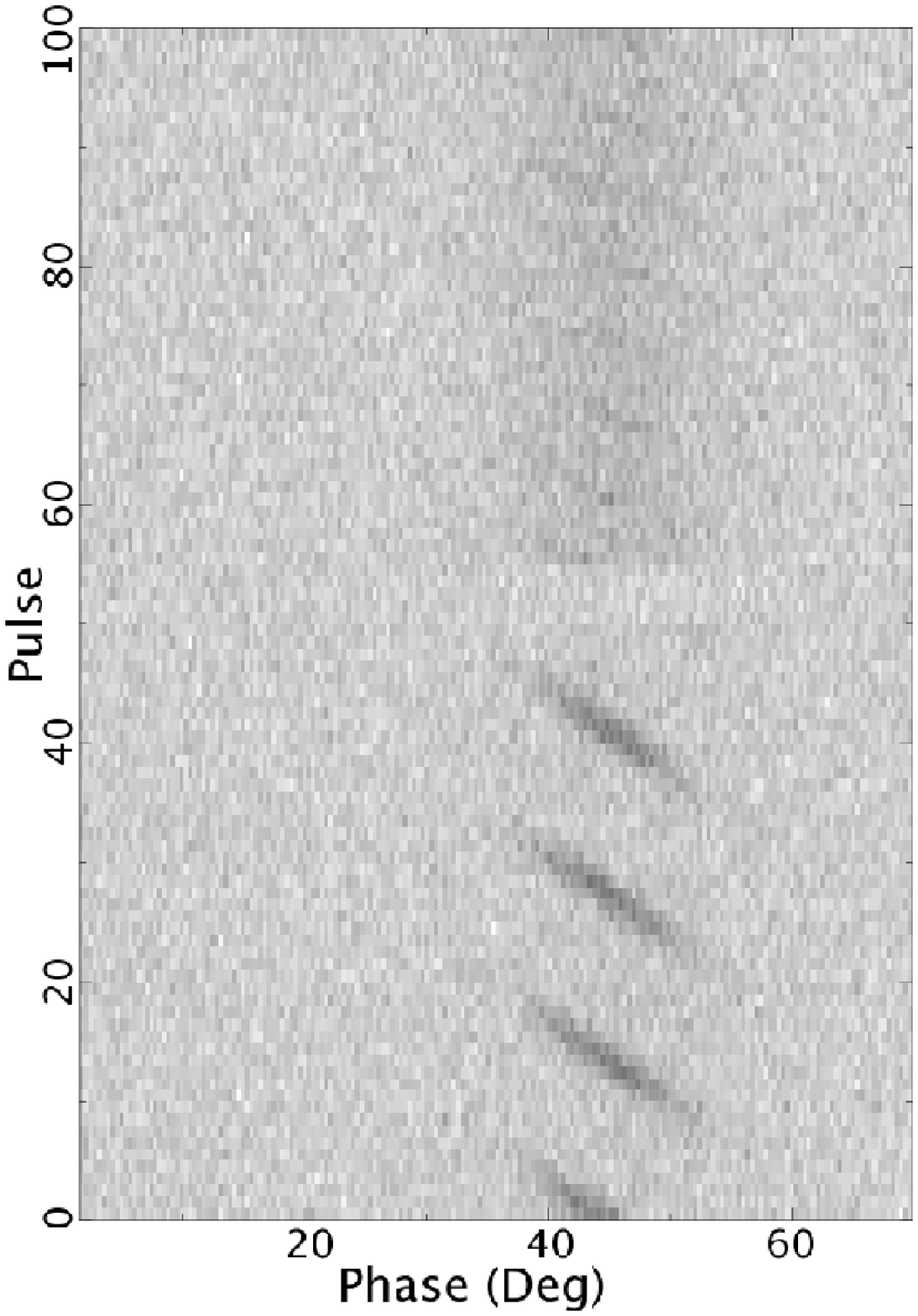} \\
  \end{tabular}

  \caption{Gray scale plots of single pulses at two frequencies from
  the simultaneous observations (left panels) and from the model
  (right panels). The upper plots are at 243\,MHz and the bottom plots
  are at 4.85\,GHz. The first 50 pulses are in drift mode A, the
  following 5 pulses are nulls and the remaining pulses are in drift
  mode B. The pulses in the model were aligned with the pulses from
  the 243-MHz observation. This results in the offset between single
  pulses from the observation and from the model at 4.85\,GHz. There is
  some interference visible in the pulses 60 to 68 of the observation
  at 4.85\,GHz.}
  \label{fig:grayscales}
\end{figure}

\begin{figure*}
  \centering
  \begin{tabular}{cc}7
  \includegraphics[width=0.25\textwidth,angle=-90]{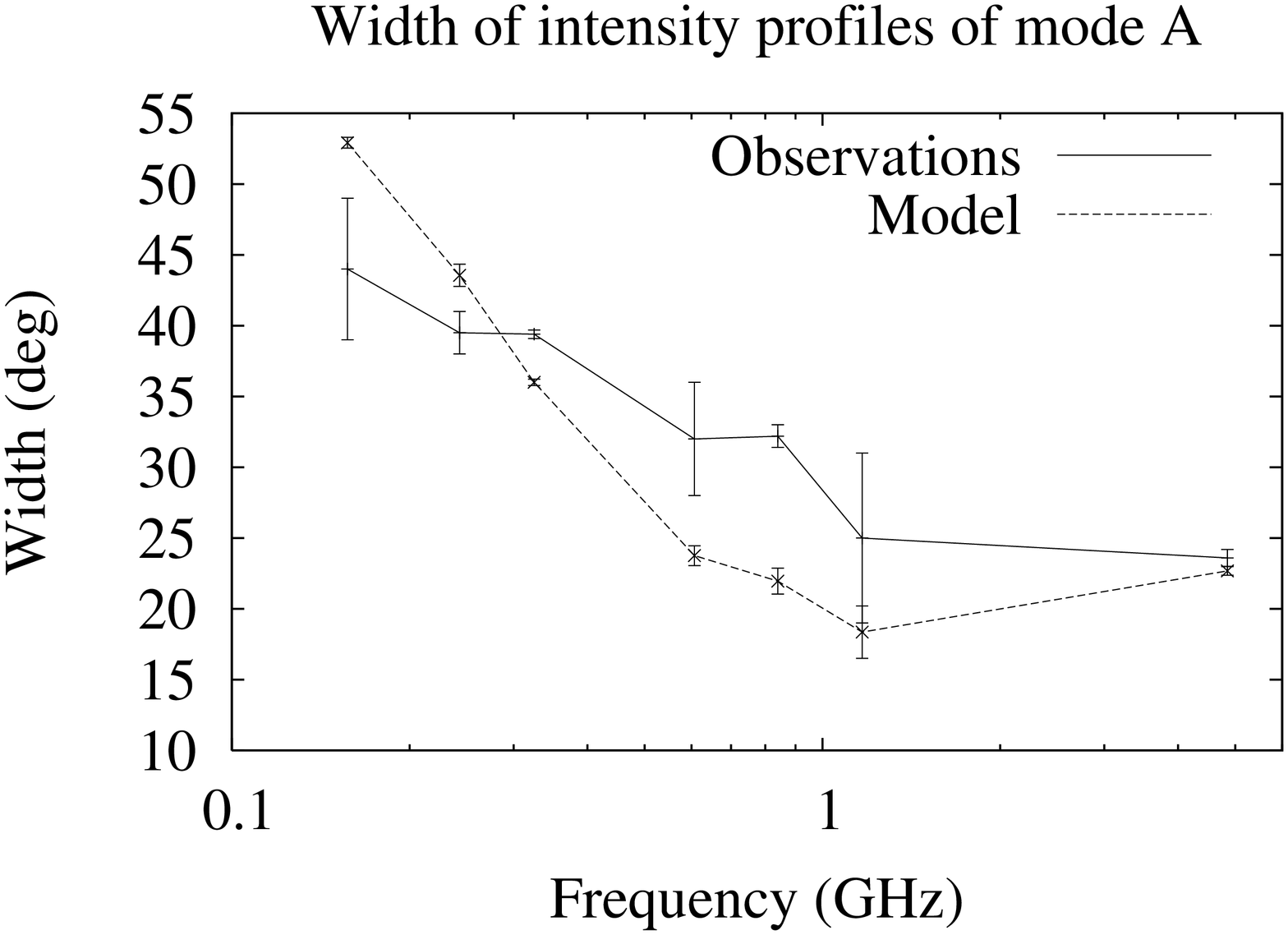} &
  \includegraphics[width=0.25\textwidth,angle=-90]{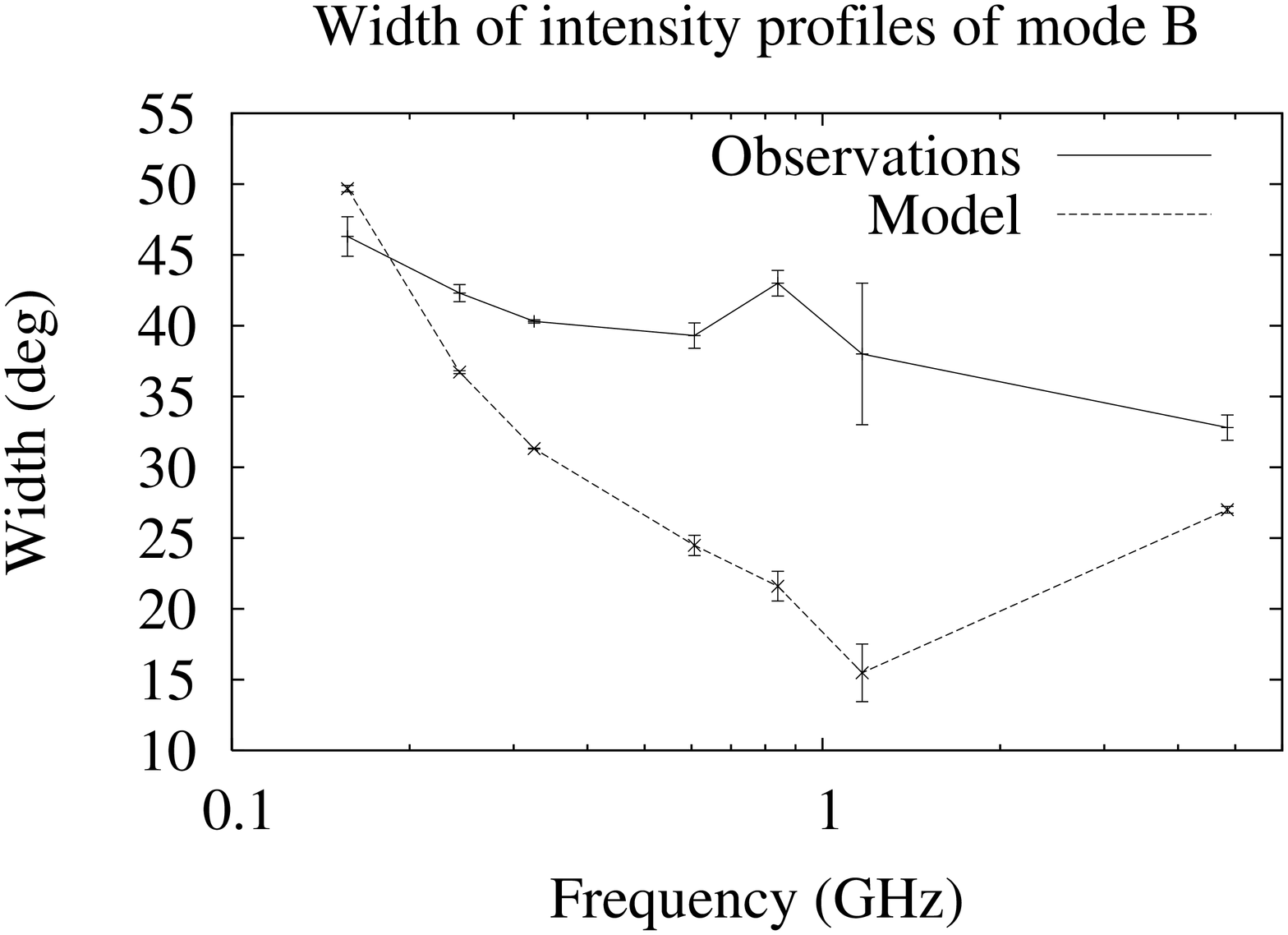}\\ [-0.2cm]
  \includegraphics[width=0.25\textwidth,angle=-90]{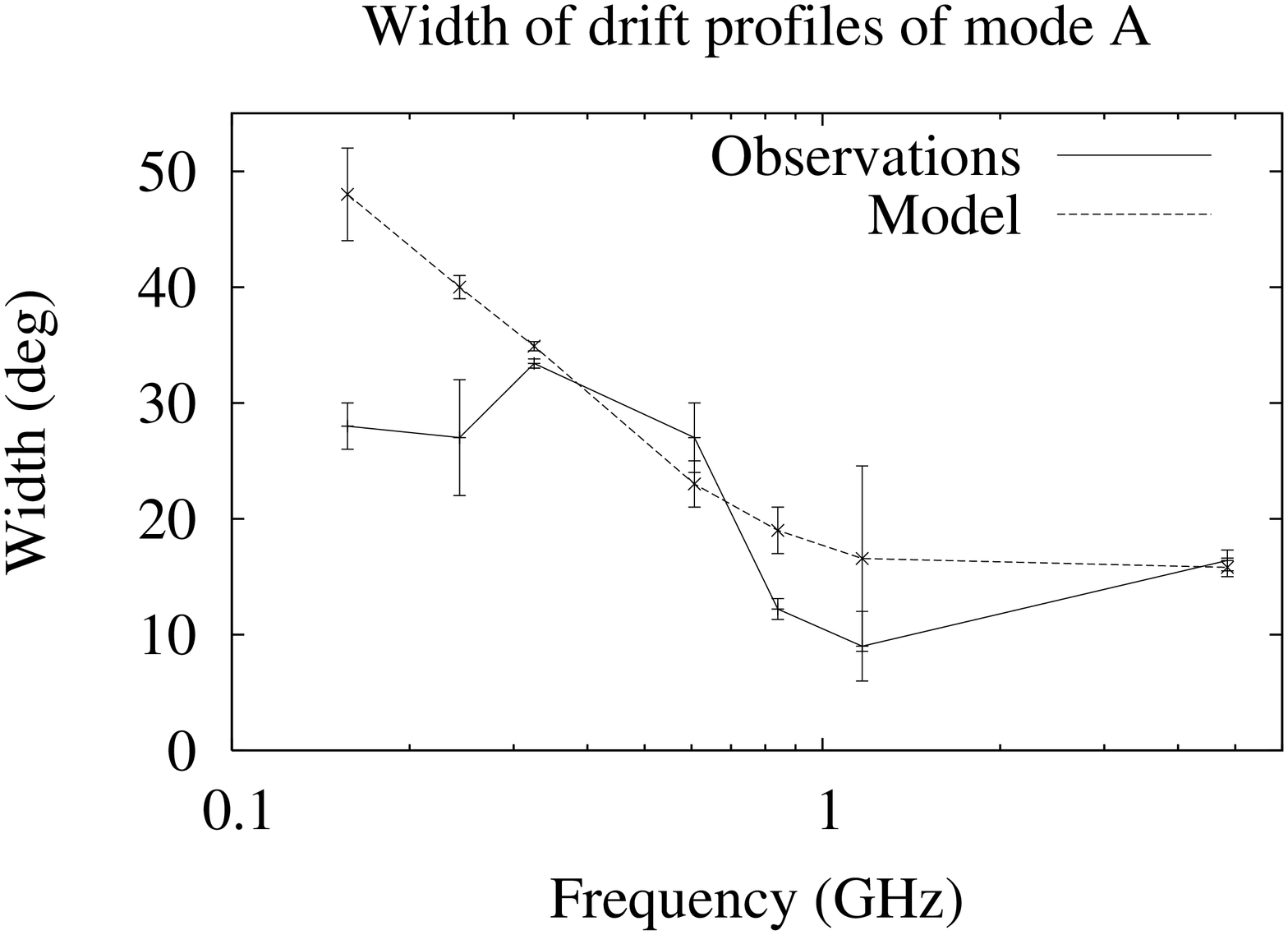} &
  \includegraphics[width=0.25\textwidth,angle=-90]{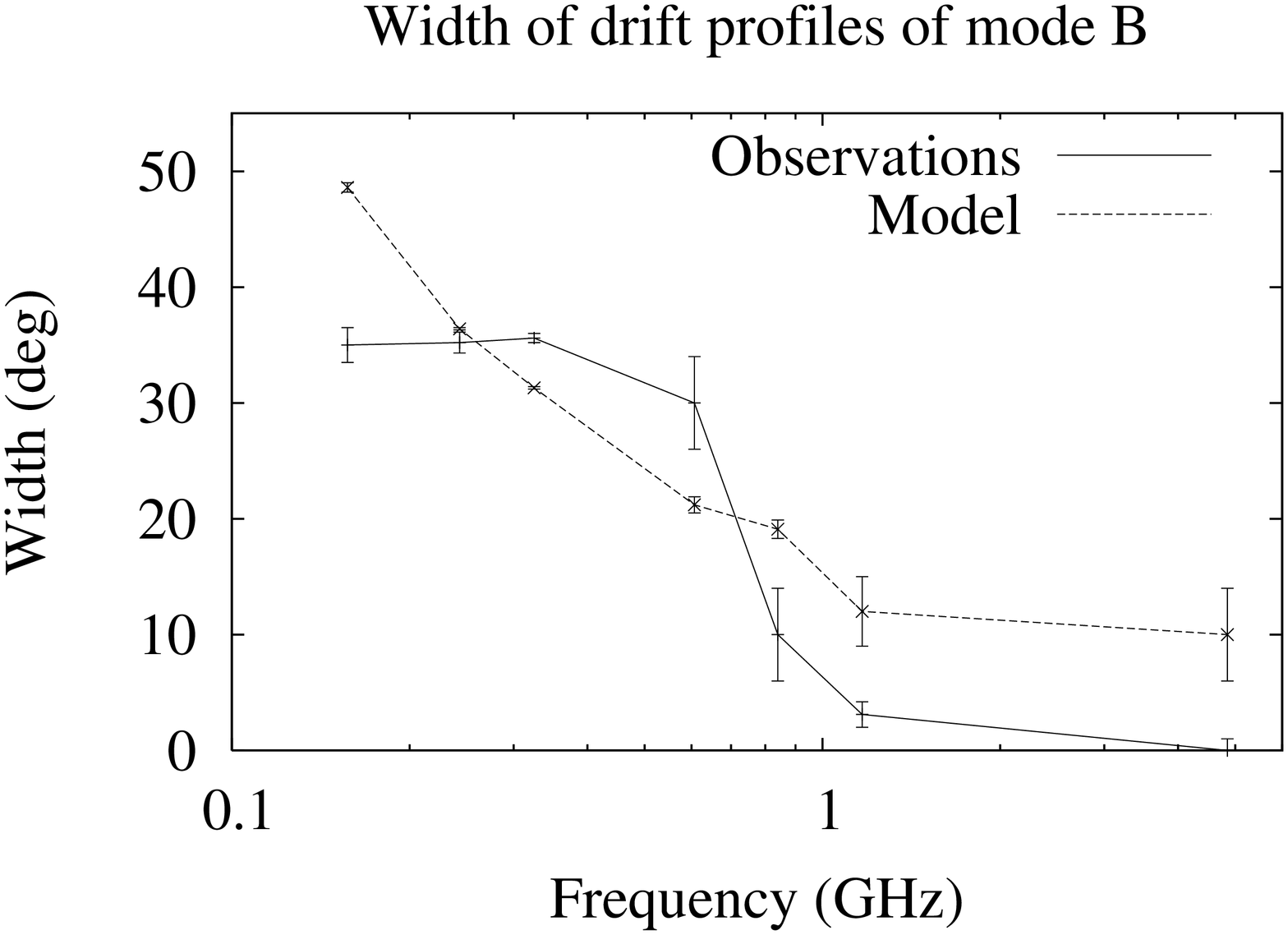}
  \end{tabular}
  \caption{Widths of the average intensity profiles (upper plots) and
  average drift intensities (lower plots) from the
  observations (solid line) and from the model (dashed line), for drift modes A
  (left plots) and B (right plots) at each frequency.}
  \label{fig:widths}
\end{figure*}

\begin{figure*}
  \centering
  \begin{tabular}{cc}
  \includegraphics[width=0.25\textwidth,angle=-90]{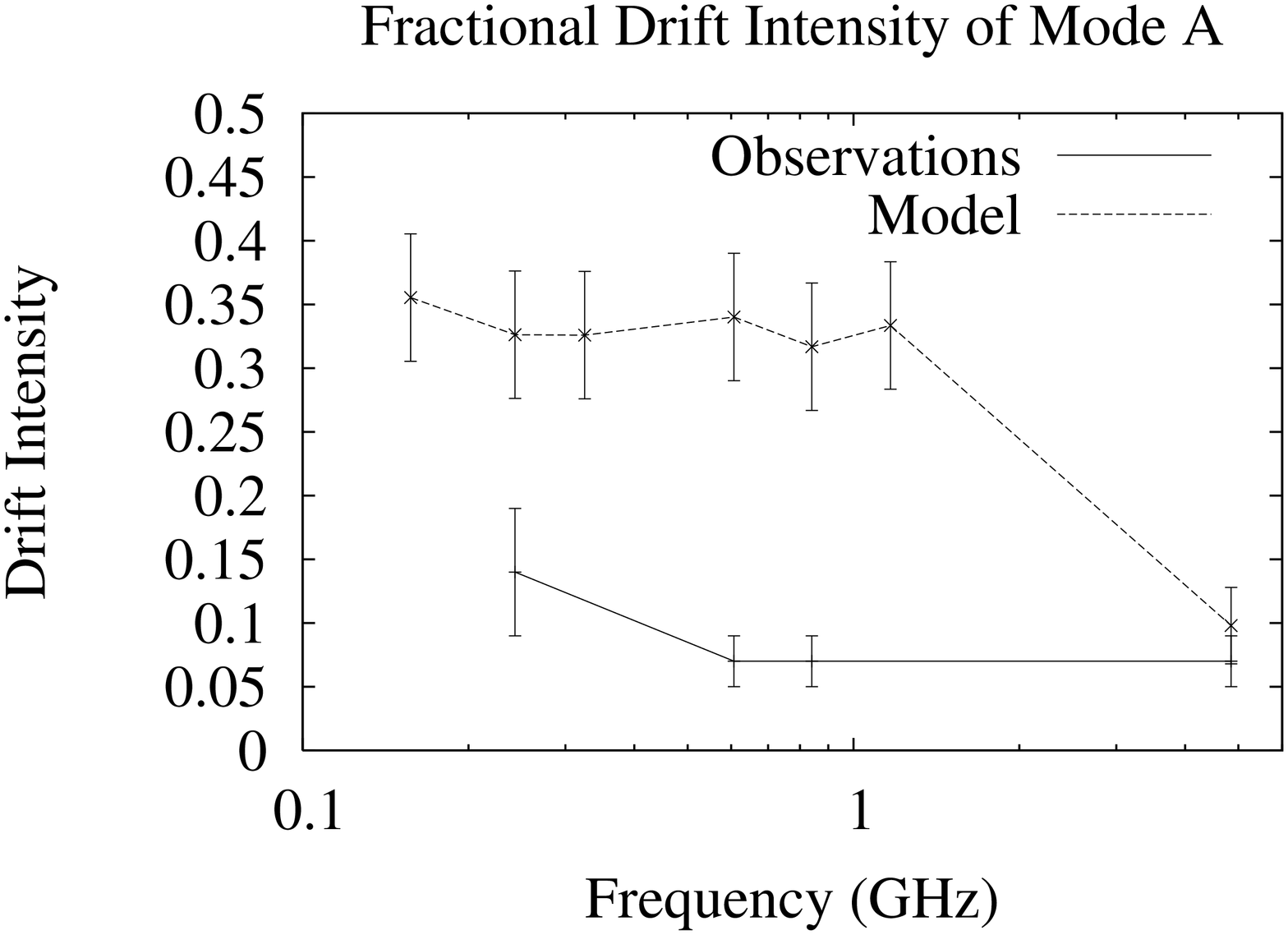} &
  \includegraphics[width=0.25\textwidth,angle=-90]{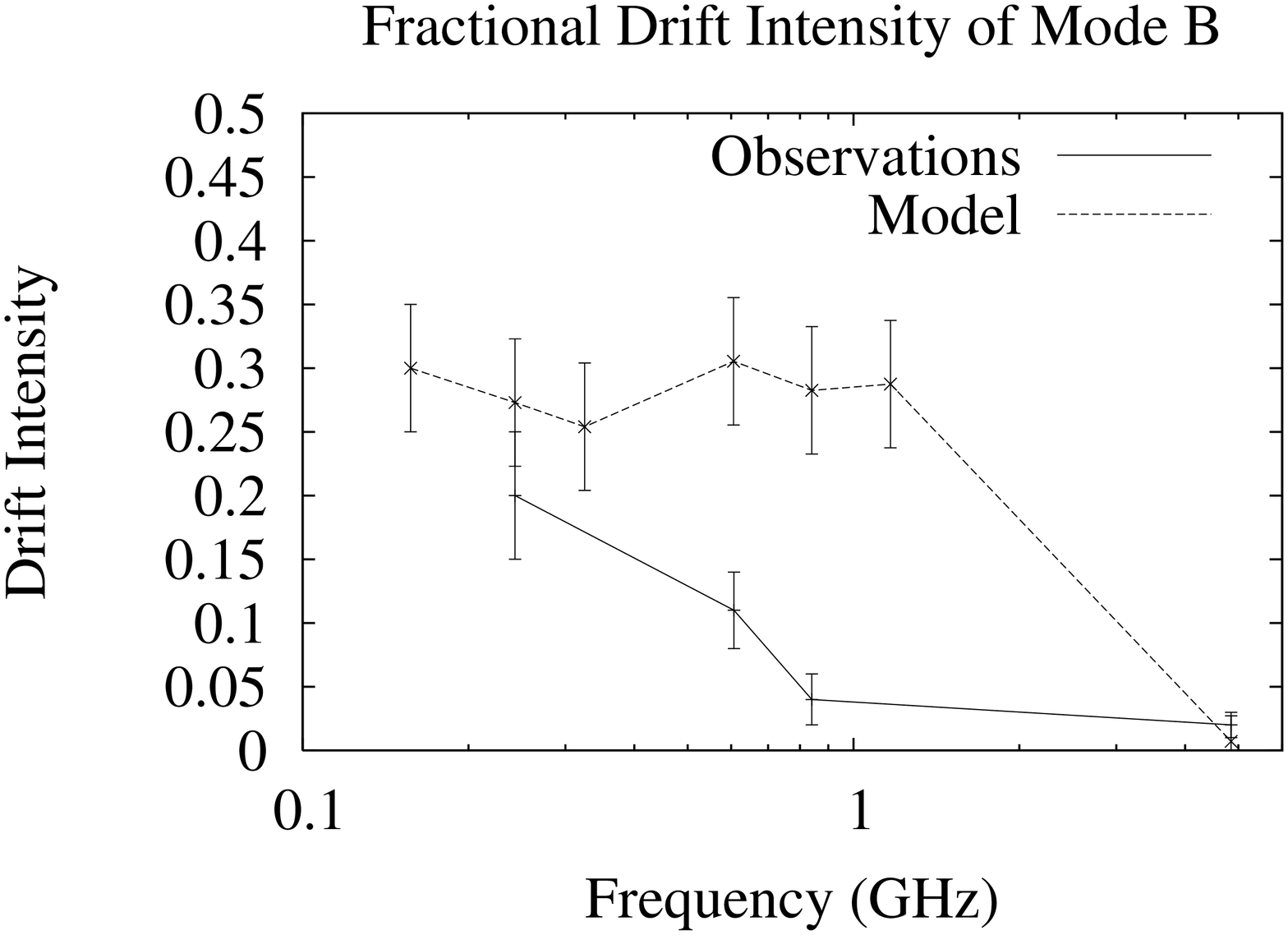}
  \end{tabular}
  \caption{Average fractional drift intensity from the observations
  (solid line) and from the model (dashed line), for drift modes A (left
  plot) and B (right plot).}
  \label{fig:fdi}
\end{figure*}

\begin{figure}
  \centering
  \includegraphics[angle=-90,width=0.35\textwidth]{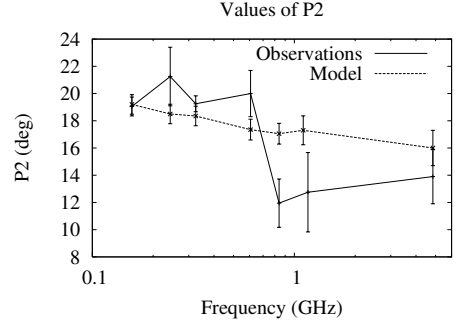}
  \caption{Value of $P_2$ from the observations (solid) and from the
  model (dashed) at each frequency. Note that the presence of
  non-drifting emission affects the method used to measure $P_2$. (See
  discussion for more details.)} 
  \label{fig:P2}
\end{figure}

\begin{figure}
  \centering
  \includegraphics[width=0.4\textwidth]{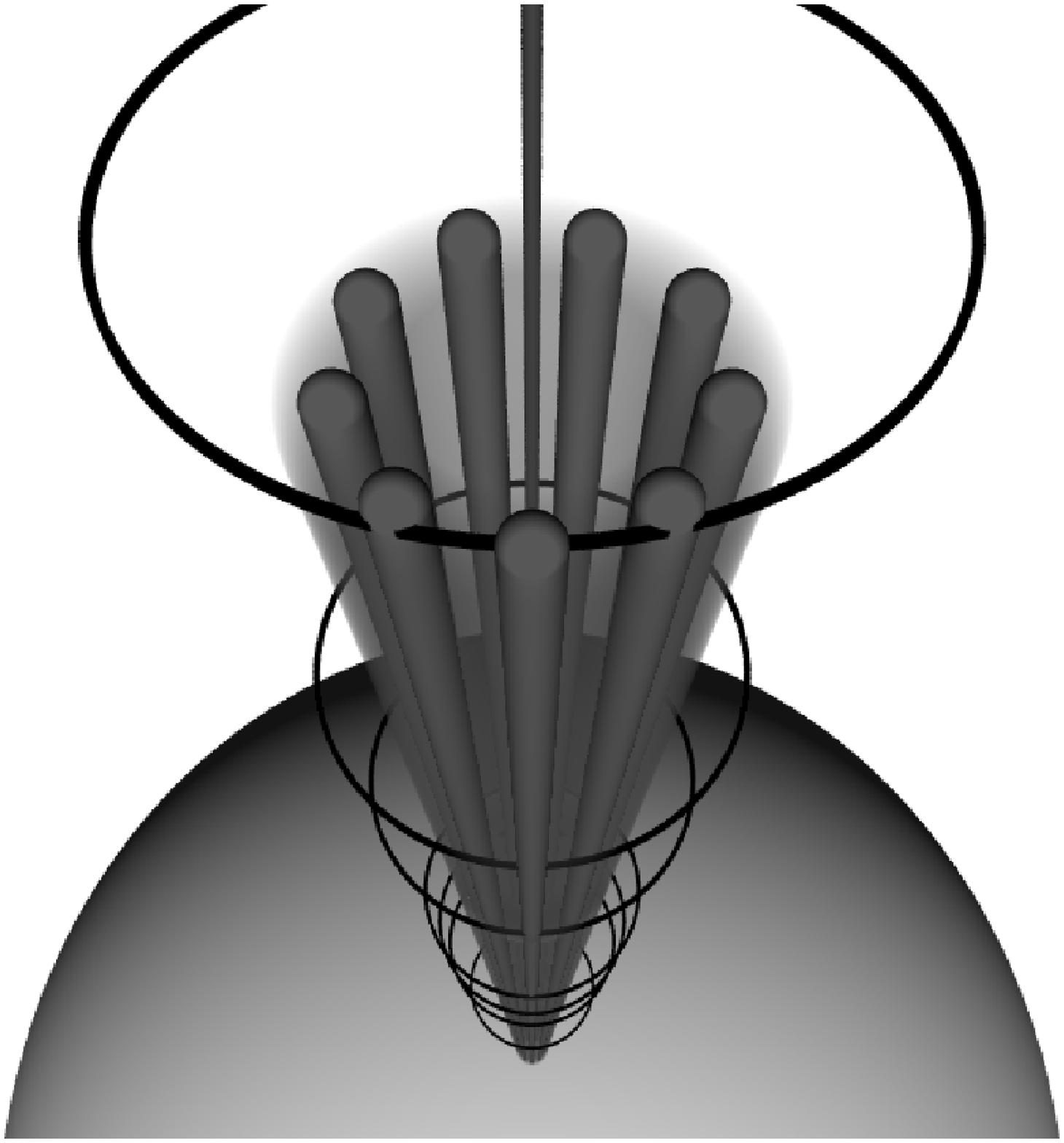}\\
  \includegraphics[width=0.4\textwidth]{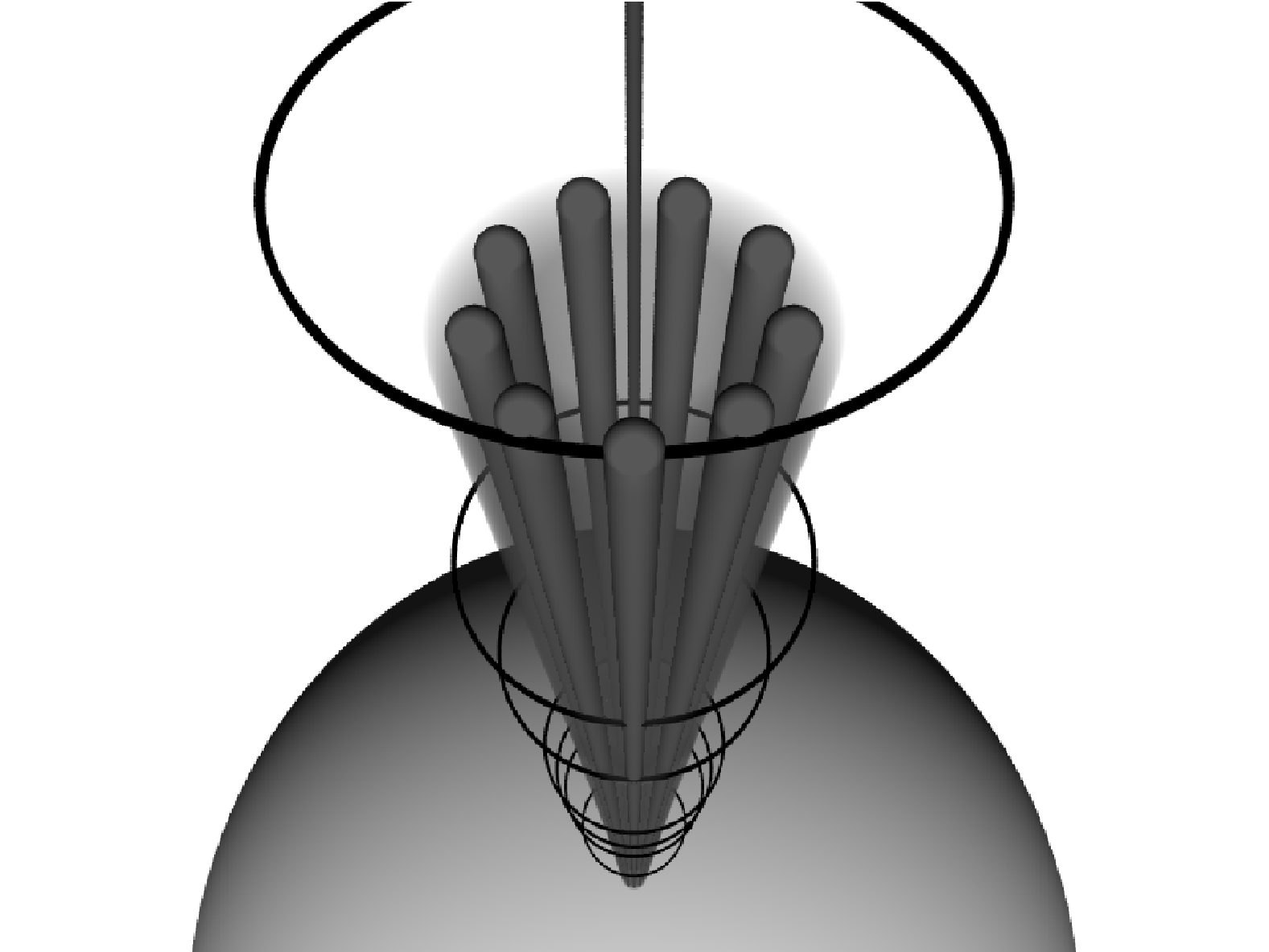}
  \caption{Two close ups of the model of the emission zone of
  \object{PSR~B0031$-$07}. The vertical line is the rotation axis of the
  pulsar. The 7 circles indicate the line of sight trajectories
  corresponding to the 7 observed frequencies. The emission zone
  consists of 9 sub-beams surrounded by diffuse emission, shown as
  semi-transparent. Both images are to scale. The top image shows the
  location of the sub-beams during mode-A drift. The bottom image
  shows the location of the sub-beams during mode-B drift, which lie
  slightly closer to the magnetic axis.}
  \label{fig:model_close}
\end{figure}

Fig.~\ref{fig:grayscales} shows a gray scale plot of the single pulses from
the observations on the left and of the single pulses from the model on the
right, for both the highest and smallest frequencies of the simultaneous
observations. It demonstrates that the model can indeed reproduce both mode A
and mode B drifts and also that the mode B drift disappears at high frequency.
It also shows the change in profile width. We further calculated the widths of
the average intensity profiles and the values for $P_2$ from the model for
both drift modes at all frequencies. For each frequency, the signal-to-noise
ratio was adjusted to be the same as the signal-to-noise ratio of the data.
The widths of both the average intensity profiles and the average drift
profiles from both the model and from the observations for drift mode A and B
at each frequency are shown in Fig.~\ref{fig:widths}. The average fractional
drift intensity from both the model and the observations for drift mode A and
B at each frequency are shown in Fig.~\ref{fig:fdi}. Fig.~\ref{fig:P2} shows
the values of $P_2$ from both the model and from the observations at each
frequency. Since the difference in $P_2$ between driftmodes due to their
different values for $\chi_0$ is far less than the error on the measured
values for $P_2$, the values of $P_2$ for drift mode A and drift mode B were
averaged together. Images of the model can be seen in
Fig.~\ref{fig:model_close}.

\begin{table}
  \centering
  \caption{Optimised parameter values in a full geometrical model for
  the emission of \object{PSR~B0031$-$07}, describing two modes of drifting
  sub-pulses superposed on low-intensity diffuse emission, assuming a
  radius-to-frequency mapping
  based on the empirical model of \citet{Thorsett91}.} 
  \begin{tabular}{ll@{\hspace{2cm}}ll}
  \hline
  \hline
  Parameter & Value & Parameter & Value \\
  \hline
  $P_1$     & 0.9429\,s         &    $\chi_0^\mro{A}$& 0.85 $\degrees$ \\  
  $R_\star$ & 10\,km            &    $\chi_0^\mro{B}$& 0.81 $\degrees$ \\ 
  $\alpha$  & 1.83$\degrees$    &    $\chi_\mro{w}$  & 0.13 $\degrees$ \\ 
  $\beta$   & 1.83$\degrees$    &    $D_\mro{w}$     & 0.22 $\degrees$ \\ 
  $h_0$     & 11\,km            &    $D_\mro{I}$     & 0.001         \\
  $\omega_0$& 1.206\,GHz        &    $N_\mro{sub}$   & 9             \\ 
  $P_3^\mro{A}$   & 13\,s       &  \\
  $P_3^\mro{B}$   & 7\,s        &  \\
  \hline
  \end{tabular}
  \label{tab:Complete_model_parameters}
\end{table}

\clearpage
\section{Discussion}
\label{sec:Discussion}
Fig.~\ref{fig:Profiles} shows that there is only one polarisation-mode
present whenever the pulsar is in drift mode A, whereas drift mode B
shows two orthogonal modes of polarisation, as has been reported by
\citet{Smits05}.  They also report that in their 325\,MHz observations
the mode-A profile shows a double component, indicating that the
geometry of the emission region during this drift mode is such that
over a few degrees of pulse phase, one sees emission from a region
between the magnetic axis and the maximally emitting field lines. This
is confirmed by the average intensity profiles from
\citet{Izvekova93}, where a double component becomes visible at
62\,MHz.  However, our current observations only show a hint of a
double component in the mode-A profiles. An explanation for this might
be that the peak of the mode-A emission zone during the present
observations lies slightly closer to the magnetic axis than the peak
of the mode-A emission zone during the observations from
\citet{Smits05}. This makes it possible that no emission was detected
from the region between the magnetic axis and the maximally emitting
field line in the low frequency observations presented here. The
position angles in Fig.~\ref{fig:Profiles} reveal that orthogonal
polarisation mode jumps occur at all frequencies. However, in some
cases the jump is not instantaneous, but rather there is a resolved
transition, as can be seen at 243\,MHz. This is not consistent with a
mode jump resulting from two competing uncorrelated orthogonal modes
of polarisation. Also, the resolved transition, present in the 607-MHz
observation is in the opposite direction of the transition that is
visible at other frequencies. This difference is not a result from an
incorrect polarisation calibration, as it can be seen that the
underlying position angle sweep is identical to that at other
frequencies. A brief study of smaller sections of the 607-MHz
observation revealed that the average position angle sweep of about
1\,000 pulses can contain an orthogonal jump in which the position
angle either increases within a few degrees of pulsar phase, decreases
within a few degrees of pulsar phase, or changes within a timescale
smaller than the timeresolution.

Further, Fig.~\ref{fig:Profiles}
shows that the average intensity profiles of \object{PSR~B0031$-$07} shift to earlier
phase with increasing frequency. This can also be seen in the average profiles
from \citet{Izvekova93}. They suggest that the shift can be removed by the
adoption of a higher value of dispersion measure. Since Izvekova et al. have
determined the dispersion measure by aligning the sub-pulses in the center of
the profile, which is similar to the method of alignment used for the present
observations, we briefly check if both the shift in their observations and the
shift in the current observations could be due to twisted magnetic field
lines. Since the sub-beams follow the magnetic field lines, the drift phase of
the sub-pulses will differ between frequencies. Thus by ``artificially''
aligning the sub-pulses between frequencies, a shift will be introduced
between the average intensity profiles. To obtain the observed shift, the
magnetic field lines need to be twisted around the magnetic axis by
approximately 20$\degrees$ between the emission altitudes observed at 157\,MHz
and 4.85\,GHz. For the present model, this implies a twist of 2$\degrees$ per
kilometer altitude. We can estimate the twist of the field lines due to a
current of charged particles with a density equal to the Goldreich-Julian
density, streaming away from the pulsar surface at the speed of light and
filling the entire polar cap. The pulse phase shift
$\Delta\phi$ over a height difference $\Delta h$ is then estimated as
\begin{equation}
\frac{\Delta\phi}{\Delta h} = \frac{B_\phi}{R_\mro{pc}B_\mro{z}} = \frac{2\pi}{P_1 c},
\end{equation}
where $B_\phi$ and $B_\mro{z}$ are the $\phi$ and z components of the
magnetic field, $R_\mro{pc}$ is the radius of the polar cap and $P_1$
is the rotation period of the pulsar. For this pulsar this leads to a
twist of only 0.0013$\degrees$ per kilometer altitude. To overcome
this difference of a factor of 1\,000, one needs to assume either
unrealistically large current densities or much greater emission
height differences.

From Fig.~\ref{fig:Driftranges} it becomes evident that
the mode-B drift is seen less frequently towards higher
frequencies. Interestingly, even though the occurrence decreases with
increasing frequency, the fractional drift intensity itself of this
drift mode does not drop until 4.85\,GHz and even at this frequency it
can still be seen occasionally. This is in contrast with the results
from \citet{Smits05} who did not see any mode-B drift
sequences in 2\,700 pulses at 4.85\,GHz, which were observed
simultaneously at 325\,MHz, nor in 5\,350 pulses at 1.41\,GHz, which
were not observed simultaneously at any other frequency. Of course, in
the absence of a simultaneous observation at low frequency, a mode-B
drift at high frequency might remain undetected. Also, due to the
infrequent occurrence of the mode-B drift at 4.85\,GHz, it is possible
that Smits et al. did not detect this drift mode at 4.85\,GHz because
it did not occur during their observation. Alternatively, the location
of the mode-B emission during the present observation might differ
from the location of the mode-B emission during the observations from \citet{Smits05}. The fact that a fractional drift
intensity can only be detected sporadically for mode B at 4.85\,GHz,
does suggest that the location of the mode-B emission can slightly
change on timescales of a few minutes. Despite these changes over
small timescales, the average fractional drift intensities that are
shown in Table~\ref{tab:driftintensity} represent how
the line of sight intersection with the emission region changes over
frequency. However, it should be noted that residual interference and
variations in the sub-pulse intensity within a drift sequence can
affect these values. These effects result in the large errors listed
in Table~\ref{tab:driftintensity}. Still, the values in
Table~\ref{tab:driftintensity} are consistent with the
suggestion that at all frequencies a clear mode-A drift can be
detected, whereas the mode-B drift becomes harder to detect towards
higher frequencies. This can also be seen in
Table~\ref{tab:width_of_driftprofiles}, where the widths
of the average mode-B drift profiles decrease with increasing
frequency. The widths of the average mode-A drift profiles can also be
seen to decrease with increasing frequency, but even at the highest
frequency, the profile is still broad enough to allow detection.

Fig.~\ref{fig:grayscales} shows that the model can reproduce the
single pulses of both drift mode A and B, except for the shift in
arrival time between the single pulses at low and high
frequencies. This shift can also be seen in the average profiles of
Fig.~\ref{fig:Profiles}.  The change of the width of the average
intensity profiles with frequency can be seen in
Fig.~\ref{fig:widths}. It is clear that the widths of the
intensity profiles from the model have a steeper spectrum than those
widths from the observations. This steep spectrum is a consequence of
the change in height by a factor of 2, which is required to obtain an
observable change in $P_2$. At 4.85\,GHz, the width from the model
becomes broader again due to the increased intensity of the broad
non-drifting emission with respect to the emission associated with the
drifting sub-pulses.  Fig.~\ref{fig:fdi} reveals that the average
fractional drift intensity from the model is overall much higher than
the average fractional drift intensity from the observations. This is
as expected, since the model does not include pulse to pulse intensity
fluctuations, which will lower the fractional drift intensity.
Alternatively, the different spectra of the drifting and non-drifting
emission might be due to a different emission process rather than a
different geometry. However, from the results presented here, we feel
that it is not necessary to assume different emission processes.  The
model does reproduce the disappearing of drift mode B at 4.85\,GHz, as
was already seen in Fig.~\ref{fig:grayscales}.  Fig.~\ref{fig:P2}
shows the change of $P_2$ over frequency from the observations and
from the model. At low frequency the change of $P_2$ from the
observation and from the model are similar. However at high
frequencies, the values of $P_2$ from the observations are
significantly lower than the values of $P_2$ from the model. A steeper
spectrum of $P_2$ in the model can be obtained by making the emission
height difference between the high and low frequencies
larger. However, this would result in a steeper spectrum for the width
of the average intensity profiles, which does not match the
observations. A possible explanation might be that the method used to
measure $P_2$ is influenced by the non-drifting emission, which we
suggest is present along with the drifting sub-pulses. The value of
$P_2$ is linearly related to the sub-pulse phase drift, which is
measured by cross-correlating consecutive pulses. In the presence of
non-drifting emission, this cross-correlation not only peaks at the
sub-pulse phase drift, but it will also have a broad peak at zero
phase shift. When the intensity of the non-drifting emission becomes
strong with respect to the intensity of the drifting emission, which
happens at high frequencies, both peaks in the average
cross-correlation will merge into one peak. This results in a lower
value for the sub-pulse phase drift and thus also for $P_2$. At low
frequencies we do not expect the value for $P_2$ to be influenced by
the non-drifting emission as it is very weak compared to the drifting
emission. And indeed, our present values of $P_2$ at low frequencies
are similar to those from \citet{Izvekova93}, or those from Fig. 9
from \citet{Bartel80}, who used auto-correlation functions to directly
measure $P_2$. Finally, Fig.~\ref{fig:model_close} show the actual
geometry and emission heights of the emission zone of \object{PSR~B0031$-$07},
as they result from the model. The lowest and highest circles
lie at heights of 2.3\,km and 13.6\,km from the surface of the star,
respectively. Fig.~\ref{fig:model_close} shows that the change in the
location of the sub-beams between mode-A and mode-B drift is very
small.

\clearpage
\section{Conclusions}
\label{sec:Conclusions}
We have shown the results from an analysis of a simultaneous multifrequency
observation of \object{PSR~B0031$-$07}. We have included two
non-simultaneous observations to obtain a total of 7 different
frequencies. From these observations we first determined the drift
mode of each drift sequence at each frequency. Contrary to what was
expected, we found multiple detections of mode-B drift at
4.85\,GHz. We then determined the position
angle sweep, the width of the average intensity profile, the width of
the average drift profile and $P_2$ for each drift mode at each
frequency. We then tried to fit three emission models to the widths of
the average intensity profiles to ultimately find a model that
describes all the observed features. We found that the models based on
curvature radiation and plasma-frequency emission could not reproduce
the frequency dependence of the widths. However, the model based on an
empirical relationship between the height and frequency of emission,
which includes an extra parameter, could reproduce the frequency
dependence of the widths for many different values of the
parameters. We therefore improved the latter model to reproduce two
drift modes of the single pulse emission.

We can summarize the features of the geometrical model of
\object{PSR~B0031$-$07} that is presented here as follows.
\begin{itemize}
\item The model reproduces to great extent the position
angle sweep (without the orthogonal mode jumps) and the frequency
dependences of the width of the average intensity profiles, the width
of the average drift profiles, the fractional drift intensity and
$P_2$, for drift modes A and B of the single pulses of
\object{PSR~B0031$-$07}. The largest deviations are found in the average
intensity profiles. 
\item The emission heights are very low. The high frequency emission
comes from a region just above the surface of the star. The low
frequency emission comes from a region about 10 kilometers higher than
the high frequency emission.
\item The parameters $\alpha$ and $\beta$ are approximately the same
and depending on the actual emission height, around 2$\degrees$ to
4$\degrees$.
\item The emission is centered around, or close to the last open field lines.
\item The emission from drift mode B comes from a region just slightly
closer to the magnetic axis than the emission from drift mode A.
\item Along with the drifting sub-pulses there is non-drifting emission
in the single pulses that becomes more significant towards higher
frequencies. This non-drifting emission might provide a hint as to why not all
pulsars show clear drifting sub-pulses.
\item Assuming that the observed drift speeds of the sub-pulses are not aliased, the number of sub-beams is around 9.
\end{itemize}

The model results in very low emission altitudes, ranging from 2.3 to 13.6\,km
above the surface of the star. This is in strong contrast with other emission
heights that have been measured for pulsars, which are typically some 10 to
1000\,km \citep[e.g.][]{Hoensbroech97b, Gangadhara01, Mitra02, Mitra04}. 
However, by improving the model, the emission heights can
become higher. The model might be improved by assuming a more realistic
particle-density distribution near the polar cap, resulting
in variable emission heights at a fixed emission frequency. This can also
improve the plasma-frequency model. Also, the emission altitude of mode A and
mode B need not be
the same, as has been assumed here.\\

\noindent\emph{Acknowledgements.} The authors would like to thank J. Rankin, G. Wright and G.
Melikidze for their rich suggestions and discussions. We also thank A. Karastergiou for his help on
data alignment and an anonymous referee for his/her usefull comments. This paper is based on
observations with the 100-m telescope of the MPIfR (Max-Planck-Institut f\"ur Radioastronomie) at
Effelsberg, the Westerbork Synthesis Radio Telescope and the Giant Metrewave Radio Telescope and we
would like to thank the technical staff and scientists who have been responsible for making these
observations possible.

\bibliographystyle{aa}
\bibliography{geometry}

\end{document}